\documentclass[12pt]{article}

\usepackage{graphics}
\usepackage{amssymb}

\newcommand{\rf}[1]{(\ref{#1})}
\newcommand{\bea}{\begin{eqnarray}}
\newcommand{\eea}{\end{eqnarray}}

\newcommand{\e}{\mbox{e}}
\renewcommand{\d}{\mbox{d}}

\renewcommand{\L}{\Lambda}
\renewcommand{\b}{\beta}
\renewcommand{\a}{\alpha}

\renewcommand{\th}{\theta}
\newcommand{\del}{\delta}

\renewcommand{\k}{\kappa}

\newcommand{\oh}{\frac{1}{2}}

\newcommand{\dg}{\dagger}

\newcommand{\tr}{\mathrm{tr}\,}
\newcommand{\ra}{\rangle}
\newcommand{\la}{\langle}

\newcommand{\mi}{\!-\!}
\newcommand{\equ}{\!=\!}
\newcommand{\pl}{\!+\!}

\newcommand{\cD}{{\cal D}}

\newcommand{\cT}{{\cal T}}

\newcommand{\cN}{{\cal N}}

\newcommand{\cG}{{\cal G}}

\newcommand{\ctN}{{\tilde{\cN}}}

\newcommand{\hT}{{\hat{T}}}
\newcommand{\hH}{{\hat{H}}}

\newcommand{\no}{\nonumber}
\newcommand{\nn}{\no\\}

\def\void{}
\def\labelmark{}

\newenvironment{formula}[1]{\def\labelname{#1}
\ifx\void\labelname\def\junk{\begin{displaymath}}
\else\def\junk{\begin{equation}\label{\labelname}}\fi\junk}%
{\ifx\void\labelname\def\junk{\end{displaymath}}
\else\def\junk{\end{equation}}\fi\junk\labelmark\def\labelname{}}

{\ifx\void\labelname\def\junk{\end{array}\end{displaymath}}
\else\def\junk{\end{array}\right.\end{equation}}
\fi\junk\labelmark\def\labelname{}\def\junk{}
}

\newcommand{\beq}{\begin{formula}}
\newcommand{\eeq}{\end{formula}}
\newcommand{\beqv}{\begin{formula}{}}

\begin{document}

\hfill AEI-2001-056

\hfill 10 June 2001

\begin{center}
\vspace{24pt}
{\Large \bf Lorentzian 3d gravity with wormholes via matrix models}

\vspace{30pt}

{\sl J. Ambj\o rn}$\,^{a},$
{\sl J. Jurkiewicz}$\,^{b}$,
{\sl R. Loll}$\,^{c}$\footnote{address from Sept `01: 
Institute for Theoretical Physics, Utrecht University, Minnaertgebouw,
Leuvenlaan 4, NL-3584 CE Utrecht} and 
{\sl G. Vernizzi}$\,^{d}$

\vspace{24pt}
{\footnotesize

$^a$~The Niels Bohr Institute, \\
Blegdamsvej 17, DK-2100 Copenhagen \O , Denmark\\
{\it email: ambjorn@nbi.dk}

\vspace{10pt}
$^b$~Marian Smoluchowski Institute of Physics, Jagellonian University,\\
Reymonta 4, PL 30-059 Krakow, Poland\\
{\it email: jjurkiew@th.if.uj.edu.pl}

\vspace{10pt}
$^c$~Albert-Einstein-Institut,\\
 Max-Planck-Institut f\"{u}r Gravitationsphysik,\\
Am M\"uhlenberg 1, D-14476 Golm, Germany\\
{\it email: loll@aei-potsdam.mpg.de}

\vspace{10pt}
$^d$~Department of Theoretical Physics, Oxford University\\
1 Keble Road, Oxford, OX1 3NP, United Kingdom\\
{\it email: vernizzi@thphys.ox.ac.uk}
}
\vspace{48pt}

\end{center}


\begin{center}
{\bf Abstract}
\end{center}
We uncover a surprising correspondence between a non-perturbative
formulation of three-dimensional Lorentzian quantum gravity
and a hermitian two-matrix model with ABAB-interaction. 
The gravitational transfer matrix 
can be expressed as the logarithm of a two-matrix integral,
and we deduce from the known structure of the latter that 
the model has two phases. 
In the phase of weak gravity, well-defined two-dimensional 
universes propagate in proper time, whereas in the
strong-coupling phase the spatial hypersurfaces disintegrate
into many components connected by wormholes.

\vspace{12pt}
\noindent


\newpage

\section{Introduction}\label{intro}

A major task of modern theoretical physics is to unite quantum
mechanics with the theory of gravity and to understand ``quantum 
geometry''. In four dimensions, this is proving a difficult task
and there is no general consensus on which direction 
to take. The so-called ``string community'', originating from 
quantum field theory, claims that M-theory provides the 
only viable road to unifying quantum mechanics and gravity, while 
people coming from the theory of relativity tend to favour approaches 
based on canonical quantization. So far neither of these 
approaches has given us a detailed understanding of the microscopic
quantum geometry of the real space-time. 

The situation is considerably better in dimension $d<4$. Although  
lower-dimensional models do not possess propagating gravitational 
degrees of freedom, their geometries are still subject to quantum 
fluctuations, and the quantum theories are non-trivial.
One may use conventional quantum field-theoretic methods to investigate
the coupling between matter and gravity, and to define and calculate
diffeomorphism-invariant correlation functions and the 
dynamically generated fractal dimension of quantum space-time.
Of course, we have no way of knowing how relevant the study of 
the lower-dimensional theories will be for the eventual theory
of four-dimensional quantum gravity, but it is clear that there are 
structural similarities and certainly some of the same questions 
can be asked.

A detailed and explicit analysis exists in two space-time dimensions.
One can calculate the anomalous dimensions acquired by matter 
fields when coupled to 2d Euclidean gravity \cite{kpz,ddk}. 
It is also understood that a typical two-dimensional Euclidean
geometry contributing to the gravitational path integral has a fractal 
structure with Hausdorff dimension four, much in the same way as a 
typical path in the path integral for a particle is also
fractal, with Hausdorff dimension two. 

The fractal structure is best understood by introducing a 
geodesic ``time'' on the two-geometries. Euclidean 2d gravity
is characterized by the fact that an infinite number of baby universes 
branches off the one-dimensional spatial slice as it evolves in 
this ``time'' \cite{kkmw,aw,ajw}. While such a process is unavoidable 
in two-dimensional gravity models coming from string theories
(describing non-critical strings), there is nothing in a theory of 
{\it quantum gravity} which demands that space should be allowed to 
split into disconnected parts. In the context of canonical gravity
this would require a ``third quantization'' to enable
the destruction and creation of baby universes, a possibility that is
not usually considered.

Not allowing for the creation of baby universes leads to 
a new two-dimensional quantum theory, called {\it Lorentzian
quantum gravity}. The name derives from the fact that the
sum over geometries in the path integral includes only a
particular subclass of Euclidean geometries which are obtained through 
a Wick rotation from a set of Lorentzian space-times with
a well-defined causal and globally hyperbolic structure \cite{al}. 
Interestingly, the resulting quantum theory 
is different from the Euclidean one. For instance, 
the fractal dimension of a typical geometry is two and not four, and 
the coupling of matter and geometry creates no anomalous 
scaling dimensions for the matter fields \cite{aal1,aal2}. The detailed 
relation between the geometries of the two models is well understood 
\cite{ackl}.

Very important in the study of two-dimensional quantum 
gravity have been methods and concepts from statistical physics 
\cite{bct,ajw}. The regularization in both the Euclidean and 
Lorentzian case uses so-called {\it dynamical triangulations}.
In this approach, geometries are created by gluing together
large numbers of identical triangular building blocks. 
The geodesic edge length of the triangles is a measure of the
fineness of the simplicial lattice and defines a 
diffeomorphism-invariant cutoff of the theory.
The action of such a piecewise linear two-geometry is calculated by 
Regge's prescription, and the state sum over geometries can in many 
cases be performed explicitly. The scaling limit (i.e. taking 
the number of triangles to infinity), defines the continuum limit 
of these models. 

Since in both Euclidean and Lorentzian 2d dynamical triangulations
we have a concept of time (``geodesic'' time and ``proper'' time
respectively), a transfer matrix can be introduced which describes
the evolution (the transition amplitude) between the spatial
configurations at time $t$ and $t+a$, where $a$ is a discrete
lattice spacing. The notion of a transfer matrix is
familiar from quantum field theories on fixed lattices.
It allows us to extract the continuum Hamiltonian of the system in
the limit as $a\rightarrow 0$ according to
\beq{1.1}
\la \phi(x)|\hat{T} |\phi'(x')\ra =   
\la \phi(x)|\e^{-a \hat{H}}|\phi'(x')\ra \to 
\la \phi(x)|(\hat{1}-a \hat{H}+O(a^2))|\phi'(x')\ra.
\eeq
This strategy has been applied successfully
in Euclidean and Lorentzian two-dimen\-sio\-nal quantum gravity.

While all of this works beautifully in dimension two,
where it has given us a number of powerful analytical tools,
the situation is quite different when one tries to use
the method of dynamical triangulations to obtain a
theory of Euclidean quantum gravity in higher dimensions.
Such models have been investigated mainly through 
numerical simulations \cite{aj,am},
and the results have so far been disappointing: no interesting
continuum limits seem to exist \cite{bielefeld}.\footnote{This seems 
to be part of a general pattern, since also other non-perturbative discrete
approaches to four-dimensional {\it Euclidean} quantum gravity 
have had little success, see \cite{livrev} for a review.} 
In fact, this was one of the main motivations for constructing
alternative {\it Lorentzian} models of dynamical triangulations,
which we have already mentioned in the two-dimensional context.
They have recently been shown to exist as well-defined regularized
models of quantum gravity also in three and four space-time 
dimensions \cite{ajl1,d3d4} (see also \cite{india} for a 
review of discrete Lorentzian gravity).

An investigation of the continuum properties of the 
three-dimensional Lorentzian gravity model has already begun. 
Computer simulations show that it avoids some of the problems of 
Euclidean simplicial quantum gravity and most likely has a continuum 
limit \cite{ajl2}, thus fulfilling some of the hopes raised
by the two-dimensional Lorentzian model. As is well-known from
previous attempts, analytic tools are hard to come by in
statistical models of quantum geometry in $d>2$. In particular,
the matrix-model methods that proved so powerful in two
dimensions have not yet been made into a useful calculational
tool in higher-dimensional quantum gravity. 
The purpose of this article is to demonstrate that in three-dimensional 
Lorentzian quantum gravity, such analytic matrix-model 
techniques {\it can} indeed be employed. 

The remainder of this article is organized as follows.
In Sec.\ \ref{model} we describe the quantum gravity model in terms
of simplicial geometries and define its transfer matrix.
At the discretized level, it is a simple variant of 
the three-dimensional Lorentzian model introduced in \cite{ajl1,d3d4} 
and studied by Monte Carlo simulations in \cite{ajl2,ajl3}. 
Motivated by some well-known properties of (2+1)-dimensional quantum
gravity, we perform an integration over all but one of
the spatial geometric degrees of freedom in Sec.\ \ref{integrate}.
In the following Sec.\ \ref{hamil} we remind the reader of
how to obtain the continuum Hamiltonian from the resulting
transfer matrix. The correspondence of 3d Lorentzian quantum
gravity with the already partially solved two-matrix model 
\cite{kz} with $ABAB$-interaction
is established in Sec.\ \ref{model}. We reinterpret the phase
structure of this matrix model in terms of geometry in
Sec.\ \ref{reinter}. This also involves a discussion 
of other, closely related matrix models with so-called 
touching-interactions.
We briefly comment on the status of the full and as yet unsolved
$ABAB$-matrix model in Sec.\ \ref{general}.
In Sec.\ \ref{cont} we describe how taking the continuum limit
in the $ABAB$-model fits in with our previous considerations of
this limit in three-dimensional gravity. We end with an outlook
in Sec.\ \ref{outlook}. The appendix contains a derivation of
the gravitational action in terms of the 3d building blocks
used in this article.

\section{Lorentzian 3d gravity from pyramids and tetrahedra}\label{model}

The motivation for constructing
non-perturbative gravitational path integrals of Lorentzian geometries
and the general properties of the dynamically triangulated
model in three dimensions were described in  
\cite{ajl1,d3d4}. For our present purposes it is convenient to 
consider a slightly modified regularization which can be related 
to a quartic matrix model. (The cubic matrix model that would 
correspond to our original model which uses only tetrahedra
has not yet been solved.) 

In the regularized model, (proper) time $t$ is discretized into integer 
lattice steps of unit one.
The spatial slices at $t \equ 0,1,\ldots $ 
are piecewise linear manifolds of spherical topology, constructed 
by gluing together flat squares with edge length $l_{s}=a$ (rather than 
the equilateral triangles with edge length $l_{s}=a$ of
references \cite{ajl1,d3d4,ajl2}).

\begin{figure}[t]
\centerline{\scalebox{0.6}{\rotatebox{0}
{\includegraphics{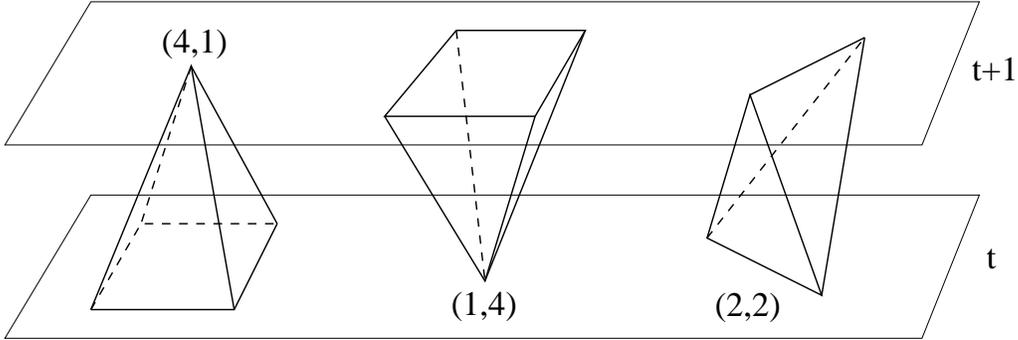}}}}
\caption[pyramids]{Pyramids and tetrahedra can be used to discretize
3d Lorentzian space-times. We show the three types of fundamental 
building blocks and their location with respect to the spatial hypersurfaces
of constant integer-$t$.}
\label{pyramids}
\end{figure}

The geometry of a spatial slice is uniquely fixed by this length 
assignment and by its connectivity matrix, specifying which pairs of
squares are glued together along a common edge. Two spatial slices
at $t$ and $t\pl 1$ form the space-like boundaries of a 
three-dimensional piecewise-linear manifold ``sandwich'' that lies in 
between the slices and has topology $S^2\times[0,1]$. 
The fundamental three-dimensional
building blocks used for ``filling in'' are regular pyramids with square
base and tetrahedra (see Fig.\ \ref{pyramids}). The pyramids have either
their base in the $t$-plane and their tip in the $t\pl 1$-plane
-- in which case we call them (4,1)-pyramids -- or vice versa for the
(1,4)-pyramids. In addition, each pyramid has four time-like edges of equal
squared length $l_{t}^{2}$ connecting the neighbouring 
slices\footnote{Our language here is ``Lorentzian'' in line with
our general philosophy \cite{ajl1,d3d4}, where a Wick rotation
corresponds to a sign flip of $l_{t}^{2}$.}. Two pyramids of equal
orientation can be glued together along a time-like triangle.
Since (4,1)-pyramids with base at $t$ and (1,4)-pyramids with base 
at $t+1$ cannot share a triangular face, we need an additional 
type of building block, namely, a (2,2)-tetrahedron with one 
space-like edge (of length $l_{s}$)
each in the $t$- and $t\pl 1$-planes (Fig.\ \ref{pyramids}). Its remaining
four time-like edges have again squared length $l_{t}^{2}$,
so that its faces can be glued to the pyramids of both types.

We will in the following use the term ``quadrangulation'' to denote
a piecewise-flat geometry made of (4,1)-, (1,4)- and
(2,2)-building blocks, as well as their two-dimensional spatial 
sections. (Note that a 3d quadrangulation may be thought of as 
a particular kind of a three-dimensional {\it tri}angulation, obtained 
by cutting all pyramids into pairs of tetrahedra.)
Starting from a $S^2$-quadrangulation at time $t=0$, we can by successive
gluing build up a three-dimensional space-time of length $t$ in the 
time direction,
consisting of $t\pl 1$ spatial spheres, and $t$ ``sandwiches'' 
in between.

Although our choice of allowed discretized space-times is of course 
motivated by the causal structure associated with the physical
Lorentzian signature, we will from now on do all calculations for
the already Wick-rotated Euclidean geometries. Without loss of 
generality, we set $l_{t}^{2}=\beta l_{s}^{2}\equiv \beta a^{2}$, for
$\beta > 1/2$ (because of triangle inequalities). Next, we must
compute the Boltzmann weight e$^{-\rm S}$ associated with a given
three-geometry. The Euclidean Einstein action in the continuum is 
given by
\beq{2.1}
S^{\rm cont} = -\frac{1}{16 \pi G_N} \int_{M} \d^3 x \sqrt{g(x)}\,(R(x)-2\L) 
 - \frac{1}{8 \pi G_N} \int_{\partial M} \d^2 x \sqrt{h(x)} \, K(x),
\eeq
where $h(x)$ is the induced metric and $K(x)$ the trace of the 
extrinsic curvature
on the boundary $\partial M$ of the manifold $M$, and $G_N$ and $\L$ 
are the gravitational and cosmological coupling constants.
As usual for 3d piecewise linear manifolds, the curvature is
concentrated on the one-dimensional edges or links, and proportional
to the deficit angle under rotation around a link in a plane 
perpendicular to it. Using the standard Regge prescription for
computing the total scalar curvature of a simplicial manifold,
the discrete counterpart of (\ref{2.1}) is derived in the
appendix, where we also discuss the inverse Wick rotation of the
action. 

Let us adopt the following notation: spatial quadrangulations at
integer-$t$ are called ${\cal T}(t)$ and a sandwich geometry
$\cG(\cT(t),\cT(t\pl 1))$. One piece of information contained in
the data characterizing the
geometry $\cG(\cT(t),\cT(t\pl 1))$ in the interval $[t,t\pl 1]$
is the set of three numbers $N_{41}(t)$, $N_{14}(t)$ and $N_{22}(t)$
of building blocks of the three types.
For a given geometry $\cG$, the action depends only on these
three bulk variables, 
\beq{2.5}
S[\cG(\cT(t),\cT(t\pl 1))] = 
c_0 - \tilde k \Big(N_{41}(t)\pl N_{14}(t)\mi N_{22}(t)\Big)+
\tilde\lambda \Big(N_{41}(t)\pl N_{14}(t)\pl  \oh N_{22}(t)\Big).
\eeq
In (\ref{2.5}), 
$N_{41}(t)\pl N_{14}(t)\mi N_{22}(t)$ is proportional to the integrated 
scalar curvature (including the extrinsic curvature terms for the boundaries)
between $t$ and $t\pl 1$, while $N_{41}(t)\pl N_{14}(t)\pl \oh N_{22}(t)$ 
is proportional to the three-volume of this piece of space-time. 
The dimensionless 
coupling constants $\tilde k$ and $\tilde\lambda$ are proportional to the 
bare inverse gravitational 
coupling constant $1/G_N$ and the bare cosmological coupling constant $\L$.
The explicit form of the coupling constants as functions of $\beta$,
$G_{N}$ and $\L$ can be found in the appendix.

When we start stacking up sandwich geometries,
the action (\ref{2.5}) is by construction additive, 
so that the total action for
a space-time extending $t$ steps in the time-direction becomes
\beq{2.5a}
S[\tilde k,\tilde\lambda]= c_0t - \tilde k (N_{41}+N_{14}-N_{22})+
\tilde\lambda (N_{41}+N_{14}+\oh N_{22}),
\eeq
where $N_{41}$, $N_{14}$ and $N_{22}$ denote now the 
total numbers of (4,1)-, (1,4)- and (2,2)-building blocks.

We can now define the transfer matrix associated with a unit step of 
proper time as\footnote{\label{footsymmetry}
Following \cite{d3d4}, we introduce quantum states $|\cT\rangle$
at fixed $t$, labelled by inequivalent {\it spatial} geometries
and normalized according to
$$
\la \cT_1|\cT_2\ra = \frac{1}{C_{\cT_1}}\, \del_{\cT_1,\cT_2},~~~~~~
\sum_{\cT} C_{\cT} |\cT\ra \la \cT| = \hat{1},
$$
where  $C_\cT$ is the symmetry factor 
of the quadrangulation $\cT$. 
} 
\beq{2.8}
\la \cT(t\pl 1)| \hT |\cT(t)\ra  = \sum_{\cG(\cT(t),\cT(t\pl 1))} 
\frac{1}{C_{\cG}} \, \e^{-S[\cG(\cT(t),\cT(t\pl 1))]},
\eeq
where the summation is over all distinct three-geometries 
$\cG(\cT(t),\cT(t\pl 1))$ whose boundary geometries 
are $\cT(t)$ and $\cT(t\pl 1)$. $C_\cG$ denotes the symmetry factor
of the quadrangulation $\cG$, i.e.\ the order of the automorphism group 
of $\cG$.

Let $\cT_{1}=\cT(t)$ contain $N_{41}(\cT(t))$ 
squares and $\cT_{2}=\cT(t\pl 1)$ contain $N_{14}(\cT(t\pl 1))$ squares. 
Substituting the sandwich action (\ref{2.5}) into (\ref{2.8}), we can write
\beq{2.8f}
\la \cT_2| \hT |\cT_1\ra  = 
\e^{-c_0}\;\e^{-(\tilde\lambda-\tilde k)(N_{41}(\cT_1)+N_{14}(\cT_2))} 
\sum_{N_{22}= N_{22}^{\min}}^{N_{22}^{\max}} 
\cN(\cT_1,\cT_2,N_{22})\; 
\e^{-(\oh \tilde\lambda +\tilde k)N_{22}},
\eeq
where the sum is over all values $N_{22}$ 
which can occur in sandwich geometries
with boundary $\cT_1 \cup \cT_2$. The combinatorial factor 
$\cN(\cT_1,\cT_2,N_{22})$ counts 
the number of distinct three-geometries (including the symmetry factor 
weights) in $[t,t\pl 1]$ for a fixed number of (2,2)-tetrahedra.

The Euclideanized amplitude for propagating a spatial geometry 
$\cT_{1}=\cT(0)$ at proper time 0 to a later geometry $\cT_{2}=\cT(t)$ 
at proper time $t$ is obtained by a
$t$-fold iteration of the transfer matrix,
\beq{2.8a}
G(\cT_{1},\cT_{2},t) =  \la \cT_{2}| \hT^t |\cT_{1}\ra,
\eeq
and satisfies the completeness relation 
\beq{2.8x}
\la \cT_{2}| \hT^t |\cT_{1}\ra = \sum_{\cT}
\la \cT_{2}| \hT^{t_1} |\cT\ra C_{\cT}\la \cT| \hT^{t_2} |\cT_{1}\ra,
\eeq
for any split $t=t_1 \pl t_2$ of the total time interval.

\section{Integrating out geometries}\label{integrate}

Let us first recall the situation in classical 3d gravity, on
space-times with compact spatial slices $\Sigma^{(g)}$ of genus $g$. 
The degrees of freedom of the theory associated with any spatial slice 
are the geometries (i.e. the spatial metrics $g_{ij}$ modulo
spatial diffeomorphisms), the elements of superspace. 
{\it A priori}, they are genuine field degrees of freedom:
for example, each metric can be decomposed uniquely (up to a
diffeomorphism) into a constant-curvature metric $\bar g_{ij}$
and a conformal factor according to $g_{ij}(x)=\e^{2\lambda (x)}
\bar g_{ij}(x)$. However, a canonical analysis reveals that the 
conformal factor $\lambda(x)$ is not a dynamical field
degree of freedom, but is completely determined by solving
the constraints. What remains is a {\it finite} number of 
so-called Teichm\"uller 
parameters (none for $g=0$, 2 for $g=1$ and $6g-6$ for $g>1$), 
coordinatizing the space of constant-curvature metrics 
for a given compact spatial manifold of genus $g$.
This implies that for the spherical case with $\Sigma^{(0)}=S^{2}$ 
not even a finite number of (classical) degrees of freedom is left after 
the conformal factor has been fixed\footnote{We mean here ``bulk'' 
degrees of freedom; depending on the choice of boundary conditions
on the initial and final spatial slices, surface degrees of
freedom may be ``liberated''.}. 

This raises the question of how these classical properties are reflected
in a gra\-vi\-tational path-integral approach of the type we are
considering. At the discretized level, we can in principle compute
the quantum amplitude between two arbitrary spatial geometries
(either at a fixed proper-time distance $t$, or for an arbitrary
distance, obtained by summing the discrete propagator (\ref{2.8a}) 
over all positive integers $t$). In general, this amplitude will
not vanish, since there are always many 3d quadrangulations 
interpolating between two given boundaries $\cT_{1}$ and $\cT_{2}$.
It implies that a generic path (i.e. a three-geometry) 
contributing to the path integral does not obey the classical 
constraints, which is not particularly surprising. However, what one 
is really interested in is the behaviour of these amplitudes in the 
continuum limit.

Previous investigations of the non-perturbative gravitational 
path integral\footnote{What is relevant for our current purposes is
always a {\it configuration} space path integral, and not one in phase 
space.} are suggestive of what may be happening in
three dimensions. The continuum analysis of \cite{dl} found that --
subject to a number of plausible conditions -- the kinetic term of
the conformal factor $\lambda$ is cancelled in the non-perturbative
path integral by a Faddeev-Popov term in the measure. This means
that there is no conformal kinetic term in the effective action,
and therefore that $\lambda(x)$ is not a propagating field degree of 
freedom. This scenario is corroborated by our numerical simulations
of 3d Lorentzian gravity \cite{ajl2,ajl3} which did not show any 
evidence of the conformal divergence closely associated
with the kinetic term for the conformal factor.

Motivated by these considerations, we will introduce a vastly
reduced set of (Hilbert space) states, by summing over all 
geometric degrees of freedom of a spatial slice of a given
two-volume $N$. Writing $| \cT_N\ra $ for a state corresponding
to a quadrangulation with $N$ squares, we define for a fixed
spatial topology the state
\beq{state}
|N\ra := \frac{1}{\sqrt{\sum_{\cT_{N}} C_{\cT_{N}}}} 
\sum_{\cT_{N}} C_{\cT_{N}} |\cT_N\ra.
\eeq
The normalization factor in front of the sum ensures the
orthonormality of the ``area states'', $\langle N | N'\rangle =
\delta_{N,N'}$. 
The discrete area $N$ may be thought of as a global conformal
degree of freedom. We do not integrate over $N$ in order to
keep control over the continuum limit and to be able to compare
our results with those of similar continuum approaches 
(see, for example, \cite{mart,hn}).
Using the new states (\ref{state}), we now make the {\it conjecture} 
that
\beq{2.13}
\la N_{14} | \hT | \cT_{N_{41}}\ra - 
\la N_{14} | \hT | \cT'_{N_{41}}\ra  \to 0 
~~~~{\rm for}~~~N_{14},N_{41} \to \infty.
\eeq
Expressed in words, this means that for large areas $N_{41}$ and
$N_{14}$ the expectation value $\la N_{14} | \hT | \cT_{N_{41}}\ra$
does not depend on which ``representative'' $\cT_{N_{41}}$ is
chosen from the set of 2d quadrangulations with $N_{41}$ squares. 
The conjecture should be understood in a probabilistic sense:
it says that the number of states 
where \rf{2.13} is {\it not} satisfied should have a slower growth
as a function of $N$ than the total number of states $|\cT_N\ra$
(which grows exponentially with $N$ for a given spatial topology).
Heuristically one can view \rf{2.13} as expressing
that in the large-$N$ limit the matrix elements 
$\la \cT_N| \hT | \cT'_{N'}\ra$ 
depend not on the two quadrangulations separately, but only 
on a suitably defined  ``distance'' between $\cT_N$ and $\cT'_{N'}$,
similar to the way the integral 
$\int\! \d x  f(x\mi y)$ over the real line is independent of $y$. 
In principle it is a combinatorial problem 
to show that \rf{2.13} is valid, but we have not yet produced such a 
proof.

An important consequence of property \rf{2.13} is that 
the completeness relation \rf{2.8x} continues to hold for the
area states $| N \ra$,
\beq{2.8i}
\la N_2 | \hT^{t_1+t_2}| N_1\ra = 
\sum_N \la N_2 | \hT^{t_2}|N\ra \la N | \hT^{t_1}|N_1\ra,
\eeq
where we have again assumed that $N_1$, $N_2$ and $N$ are all large.
We therefore still have a transfer matrix formalism, but with
the transfer matrix $\hT$ acting only 
on the subspace spanned by the linear combinations $\{ |N\ra\}$
of the original Hilbert space. Obviously, the task of diagonalizing
the transfer matrix on this reduced space is considerably simplified
compared with the original problem. 

From now on, we will focus our interest on solving the combinatorics
of a single sandwich geometry. We can rewrite relation \rf{2.8f} in an 
obvious notation as
\beq{2.8u}
\la N_{14}|\hT | N_{41}\ra = 
\e^{-c_0-(\tilde\lambda -\tilde k)(N_{41}+N_{14})} 
\sum_{N_{22}= N_{22}^{\min}}^{N_{22}^{\max}} 
\cN(N_{41},N_{14},N_{22})\; 
\e^{-(\oh \tilde\lambda +\tilde k)N_{22}},
\eeq 
where 
\beq{2.8v}
\cN(N_{41},N_{14},N_{22}) = \sum_{\cT_{N_{41}},\cT_{N_{14}}}
\cN(\cT_{N_{41}},\cT_{N_{14}},N_{22})
\eeq
denotes the total number of quadrangulations of the space-time
between $t$ and $t\pl 1$, including sums over the connectivities 
of the spatial boundary geometries at $t$ and $t\pl 1$. 
One may think of \rf{2.8v} as describing the combinatorics of
quadrangulating a sandwich geometry with {\it free} boundary conditions, 
except for the areas of the two boundaries which are kept fixed.

In order to simplify the combinatorics further, we 
introduce {\it boundary cosmological constants} $\L_i$ and 
$\L_f$ associated with the initial and final boundaries at 
$t$ and $t\pl 1$. 
They do not have an immediate physical interpretation and should
simply be thought of as convenient book-keeping devices that will be set
to zero at the end\footnote{This is analogous to the introduction of
external sources into a quantum field-theoretic path integral in
order to obtain Green's functions through functional differentiation.}.
They give rise to an additional term
\beq{2.8b}
\Delta S = \L_i a_s^2 N_{41}+ \L_fa^2_s N_{14} \equiv z_i N_{41}+z_fN_{14}
\eeq
in the action \rf{2.5}, which will allow us to introduce an asymmetry 
in the coupling constants multiplying the two areas
$N_{41}$ and $N_{14}$. 
We can use the dimensionless boundary cosmological terms 
$z_{i}$ and $z_{f}$ to obtain
the Laplace transform of the transfer matrix 
with respect to both the initial and final areas, $N_{i}\equiv N_{41}$ and 
$N_{f}\equiv N_{14}$, namely,
\beq{2.8r}
\la z_f| \hT | z_i\ra = \sum_{N_i,N_f} 
\e^{-z_i N_i-z_fN_f} \la N_f | \hT | N_i\ra.
\eeq
The matrix elements $\la N_f | \hT | N_i\ra$ are calculated for 
$z_i \equ z_f \equ 0$. From a combinatorial point of view, 
$\la N_f | \hT | N_i\ra$ counts the three-geometries $\cG$ for given 
numbers $N_i$ and $N_f$, each with relative weight $e^{-(\oh 
\tilde\lambda+\tilde k)N_{22}}$,
and $\la z_f| \hT | z_i\ra $ plays the role of a generating function for 
these numbers. 
It is usually much easier in combinatorial problems 
to calculate the generating function rather than the actual numbers.
This is well illustrated by both
Lorentzian and Euclidean simplicial quantum 
gravity in dimension two, where boundary 
cosmological constants are introduced in an analogous manner, and
where the associated generating functional greatly simplifies 
the counting of geometries. 

The reason for the simplification is the fact that by going
to $\la z_f| \hT | z_i\ra$ we have achieved
totally free boundary conditions, since the constraints of fixed 
areas $N_i$ and $N_f$ have been lifted. On the other hand, 
no information has been
lost, since we can in principle always rederive 
$\la N_f | \hT | N_i\ra$ by an inverse Laplace transformation
from $\la z_f| \hT | z_i\ra$. Given the one-step propagator
$\la N_f | \hT | N_i\ra$, we can finally obtain the propagator 
$G(N_i,N_f,t)$ for arbitrary times by iterating according
to \rf{2.8i}.

\section{Extracting the Hamiltonian}\label{hamil}

Once the matrix elements $\la N_f | \hT | N_i\ra$ are known,
one may try to extract the continuum Hamiltonian operator 
$\hH$ of the system by expanding them in the lattice spacing $a$ and
then taking $a \to 0$,
\beq{2.8w}
\la N_f | \hT | N_i\ra = \la N_f | \e^{-a \hH} | N_i\ra = 
\la N_f | \Big(\hat 1 -a \hH + O(a^2)\Big)| N_i\ra.
\eeq  
In this way one obtains the quantum Hamiltonian
in the ``$N$-representation''. It is also possible (and usually 
easier) to extract $\hH$ from the Laplace transform of the 
one-step propagator, $\la z_f| \hT | z_i\ra$ (yielding the
Hamiltonian in the ``dual'' $z$-representation).  

Let us illustrate this by a concrete
calculation in 2d Lorentzian gravity, where the space-time has
topology $S^{1}\times [0,1]$. In this case, the
generating functional $\la z_f| \hT | z_i\ra$ is known 
explicitly\footnote{In line with our construction in dimension 
three, we consider here the symmetric propagator of \cite{al} with
unmarked boundary loops.} \cite{al},
\beq{2d.1}
\la z_f| \hT | z_i\ra = 
\log \left(\frac{(1-\e^{-(\lambda_{2}+z_i)})(1-\e^{-(\lambda_{2}+z_f)})}{1-
\e^{-(\lambda_{2} +z_i)}-\e^{-(\lambda_{2}+z_f)}}\right).
\eeq
The relevant bare couplings in two dimensions are the bulk cosmological
constant $\lambda_{2}$ and the boundary cosmological constants
$z_i$ and $z_f$. They are related to the renormalized continuum
coupling constants $\L_{2}$, $\L_{i}$, $\L_{f}$ by
\beq{2d.2}
\lambda_{2} = \log 2+\oh \L_{2} a^2,~~~~z_{i,f} = \L_{i,f} a.
\eeq
It is straightforward to expand \rf{2d.1} to lowest non-trivial 
order in the lattice spacing $a$ and thus obtain the matrix elements
$\la \L_f | \hH | \L_i \ra$. We have
\beq{2d.s1}
\la \L_f| \hT | \L_i\ra  = {\rm analytic}-\log (\L_i+\L_f) - 
a \frac{\L_{2}-\oh(\L_i^2+\L_f^2)}{\L_i+\L_f} +O(a^2),
\eeq
where ``analytic'' refers to a constant term and to 
terms linear in $\L_i$ and $\L_f$.
By an inverse Laplace transformation we can change variables 
from the boundary cosmological constants $\L_{i,f}$ to the 
conjugate (continuum) length variables $L_{i,f}$ of the boundaries.
To first order in $a$, one finds 
\bea
\la L_f | \hT | L_i \ra &=& 
\int_{-i\infty}^{i \infty} \frac{\d \L_i}{2\pi i} 
\int_{-i\infty}^{i \infty} \frac{\d \L_f}{2\pi i}
\; \e^{\L_i L_i+\L_f L_f} \; 
\la \L_f| \hT | \L_i\ra \label{2d.s2}\\
&=& \frac{1}{L_i}\del(L_i-L_f)-a\left(-\frac{\d^2}{\d L_f^2}+\L_{2} \right)
\del(L_i-L_f), \no
\eea
where we have ignored the analytic terms which lead to non-propagating 
terms of the form $\del(L_f) \del(L_i)$. The appearance of such
non-universal terms is familiar from the transfer matrix of
2d Euclidean gravity and can be accounted for by a careful treatment
of the boundary conditions at $L=0$ \cite{aw,watabiki,aknt}. 

In complete analogy with the definition given in footnote 
\ref{footsymmetry},  
the symmetry factor of a spatial $S^{1}$-boundary of length $L$ is 
given by $C_{L}=1/L$
and there is an orthogonal basis $\{ |L\rangle\}$ of the continuum 
Hilbert space, obeying the normalization conditions
\beq{normal}
\la L_1 | L_2\ra = \frac{1}{L_1} \del(L_1-L_2), ~~~~~
\int_0^\infty \d L \, |L\ra L \la L| = \hat{1}.
\eeq
We deduce that the Hamiltonian operator in the ``$L$-representation'', 
acting on functions $\psi(L) = \la L|\psi\ra$, is   
\beq{2d.s3}
\hH =- \frac{\d^2}{\d L^2}L + \L_{2} L,
\eeq
which is hermitian with respect to the measure $L \d L$, as it should 
be.

In principle we would like to use the same strategy to determine 
the quantum Hamiltonian $\hH (A)$ and its spectrum also 
in the three-dimensional case. Of course we cannot be sure 
that this will lead to a simple differential operator as a function of
the spatial volume, as was the case in $d=2$, \rf{2d.s3}. 
Even in two dimensions, there are models inspired by 
Lorentzian gravity where this is not the case \cite{charlotte}. However,
in the 2d cases studied so far it has always been possible to turn 
$\hH$ into a local differential operator by transforming to a variable 
different from $L$.

\section{The matrix model correspondence}\label{matrix}

We will now relate the Laplace-transformed one-step propagator 
\rf{2.8r} to a matrix model. This will be done by
showing that the two-dimensional configurations associated with a slicing
at {\it half}-integer $t$ of a sandwich geometry appear as terms in the 
perturbative expansion of a hermitian two-matrix model with 
$ABAB$-interaction.
(A similar observation for the original Lorentzian 3d simplicial 
model was already made in \cite{ajl2}.)

Imagine a one-step geometry $\cG$ obtained by gluing the three
types of building blocks of Fig.\ \ref{pyramids}. The intersection of this
three-geometry with the spherical constant-time hypersurface at 
$t\pl \oh$ can be visualized as a pattern
of squares, whose edges correspond to the intersections of the
time-like triangular faces of the 3d building blocks with this surface.
Let us distinguish the two cases where the time-like
triangle has its base either in the $t$- or the $t+1$-quadrangulation
by drawing its intersection at $t\pl \oh$ as either a solid or
a dashed line. Thus a (4,1)-pyramid gives rise to a square of solid 
edges, a (1,4)-pyramid to one of dashed edges and a (2,2)-tetrahedron to 
a square with alternating solid and dashed edges. The way in which
these two-dimensional building blocks appear in the 
$S^2$-quadrangulation at $t\pl \oh$ is that they can be glued to 
each other only along pairs of edges of the same type.

\begin{figure}[t]
\centerline{\scalebox{0.6}{\rotatebox{0}
{\includegraphics{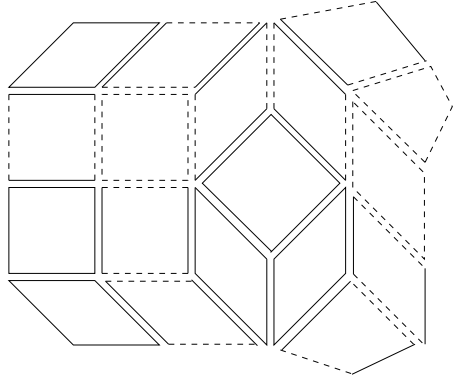}}}}
\caption[quadran]{A piece of a typical quadrangulation at 
$t+1/2$. The three types of squares made from solid and dashed lines
arise as sections of the (4,1)-, (1,4)- and (2,2)-building blocks.}
\label{quadran}
\end{figure}
\begin{figure}[t]
\centerline{\scalebox{0.6}{\rotatebox{0}
{\includegraphics{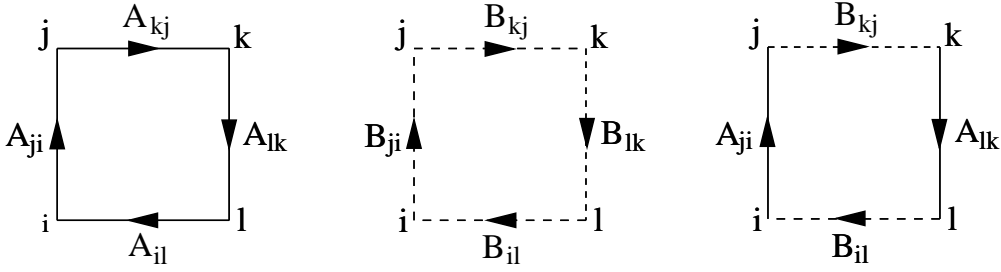}}}}
\caption[glue]{Matrix-model representation of the building blocks
at $t+1/2$. The gluing rules for the squares are determined by the
Gaussian integrations, $\langle A_{ij}A_{kl}\rangle =\delta_{il}
\delta_{jk}$, $\langle B_{ij}B_{kl}\rangle =\delta_{il}
\delta_{jk}$, and $\langle A_{ij}B_{kl}\rangle =0$.}
\label{glue}
\end{figure}

The quadrangulation at $t\pl \oh$ can thus be viewed as a double-line graph
of the kind illustrated in Fig.\ \ref{quadran}. This type of graph is 
generated in the large-$M$ limit by the two-matrix model
\beq{2.9}
Z(\a_1,\a_2,\b) = \int \d A \d B \, 
\e^{-M \tr ( \oh A^2+\oh B^2 -\frac{\a_1}{4} A^4
-\frac{\a_2}{4} B^4 -\frac{\b}{2} ABAB)},
\eeq
where $A$ and $B$ are Hermitian $M\times M$ matrices. By expanding the 
non-Gaussian part of the exponential in powers of $\a_1$, $\a_2$ and $\b$
and performing the Gaussian integral we are led by Wick's theorem to 
a successive gluing of the three kinds of squares described 
above if we make the index assignments as shown in Fig.\ \ref{glue}. 

As usual, the logarithm of the partition function of the model,
$ M^2 F(\a_1,\a_2,\b) \equ \log Z(\a_1,\a_2,\b)$, generates only
connected quadrangulations, and taking the 
large-$M$ limit will select those with $S^{2}$-topology. In 
principle $F$ may be expanded in a power series in $M^{-2}$, 
with higher-order contributions corresponding 
to quadrangulations of higher genera. 
Although we are mainly interested in the spherical limit,
anything we say could be repeated for higher-genus surfaces, 
thus relating to 3d gravity on space-times with topology 
$\Sigma^{(g)}\times [0,1]$.
We can now write out the generating function $F$ as an explicit power 
series,
\beq{2.10}
F(\a_1,\a_2,\b)=\sum_{N_{41},N_{14},N_{22}} 
\ctN(N_{41},N_{14},N_{22}) \; \a_1^{N_{41}}\a_2^{N_{14}}\b^{N_{22}},
\eeq
where $\ctN(N_{41},N_{14},N_{22})$ 
denotes the number of (connected) spherical quadrangulations
described above, including symmetry factors. 

Comparing the form of \rf{2.10} with the previous expressions
\rf{2.8u} and \rf{2.8r}, and making the identifications\footnote{
\label{footvanish}Adopting \rf{2.12}, the limit of vanishing 
boundary cosmological constants $z_{i,f}$ corresponds to setting
$\a_1 \equ \a_2$. Unfortunately we cannot put $\a_1 \equ \a_2$ before 
having extracted the matrix elements $\la N_f|\hT | N_i\ra$ from
$F(\a_1,\a_2,\b)$.} 
\beq{2.12}
\a_1 = \e^{\tilde k-\tilde\lambda -z_i},~~~ 
\a_2 = \e^{\tilde k-\tilde\lambda -z_f},~~~
\b = \e^{-(\oh \tilde\lambda +\tilde k)},
\eeq
one could be tempted to conclude that 
\beq{2.11}
F(\a_1,\a_2,\b) \stackrel{?}{=} \la z_f|\hT | z_i \ra.
\eeq
However, this is {\it not} correct, since the number of configurations
generated by the matrix model is strictly larger than those obtained 
from the Lorentzian gravity model, that is,
\beq{ineq}
\ctN(N_{41},N_{14},N_{22}) > \cN(N_{41},N_{14},N_{22}).
\eeq
This difference is intimately connected to the fact that 
in the gravity case, the allowed configurations are not so much
two-dimensional structures {\it per se} as sections of
larger, three-dimensional objects whose three-dimensional
manifold structure is encoded in the colouring (the dashed and
solid lines) of the two-dimensional graph.

\begin{figure}[t]
\centerline{\scalebox{0.6}{\rotatebox{0}
{\includegraphics{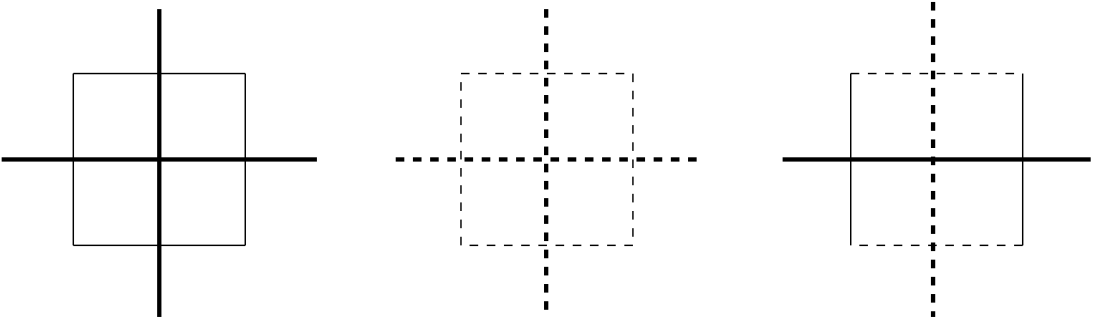}}}}
\caption[dual]{Vertices of $\phi^4$-graphs dual to spatial
quadrangulations.}
\label{dual}
\end{figure}

The generalization inherent in the matrix model is best described
in terms of the graphs dual (in a two-dimensional sense) 
to the quadrangulations. In this more conventional picture, the terms 
$\tr A^4$, $\tr B^4$ and $\tr ABAB$ are represented by four-valent
vertices (placed at the centres of the squares of the original 
2d quadrangulation) 
with four outgoing solid lines, four dashed lines or alternating
solid-dashed lines (see Fig.\ \ref{dual}). The labels $A$ and $B$ 
are now associated with the dual solid and dashed edges connecting pairs 
of such vertices. 

In order to discuss the regularity conditions that must be
satisfied by such a dual graph to qualify as (the dual of) a section of
a Lorentzian three-geometry, we define a $A$-loop (a $B$-loop) as a
closed sequence of solid (dashed) dual edges with no further solid
(dashed) dual links in its interior. This interior region (which has the
form of a two-dimensional disc whose boundary is the loop) we call
a $A$-domain (a $B$-domain). A dual graph coming from a 3d Lorentzian
geometry then satisfies the following constraints (c.f. App.\ 2 of
\cite{ajl2}):
\begin{itemize}
\item[(1)]
The two subgraphs formed from only $A$-edges and only $B$-edges 
must each be connected. 
\item[(2)]
The two separate $A$- and $B$-subgraphs can have neither tadpoles nor 
self-energy subdiagrams. This ensures that they are associated
with regular 2d simplicial manifolds at the times $t$ and 
$t+1$ respectively. 
\item[(3)]
The intersection of any pair of $A$- and $B$-domains cannot be multiply 
connected. This implies that any pair of vertices, one at time $t$ and
one at time $t+1$, of the original quadrangulation cannot be connected 
by more than one time-like link.
\item[(4)]
The (one-dimensional) intersection of a $A$-loop with a $B$-domain
(and vice versa) must be either empty or simply connected. 
This implies that any three vertices of the original quadrangulation
cannot belong to more than one time-like triangle.
\end{itemize}
Thus we see that the dual graphs coming from the matrix model are
considerably more general than those associated with 3d simplicial
space-times. The generalizations occur in several ways. The
matrix-model graphs can 
have arbitrary numbers of disconnected spherical $A$- and $B$-subgraphs
(the only requirement being that the combined graph is spherical and 
connected) and each of these may contain
(generalized) self-energy diagrams and tadpoles. Furthermore,
the $A$- and $B$-loops can freely meander around each other, with
arbitrary numbers of mutual intersections. 

We conjecture that the conditions (2)-(4) are not 
important in the sense that their implementation or otherwise will
not affect the continuum properties of the model. We think that
they constitute merely a 3d generalization of the universal behaviour
already observed for 2d matrix models. In that case the inclusion
of tadpole and self-energy subgraphs leaves the continuum limit
unchanged (although it implies an enlargement of the configuration
space from genuine 2d simplicial manifolds to more general 2d
combinatorial complexes).

We have checked numerically that the same happens in our 3d model.
More precisely, we have performed simulations for
the original simplicial model with tetrahedral building blocks of 
\cite{ajl1,ajl2,d3d4} where one can formulate conditions completely
analogous to (1)-(4) above. Dropping then the constraints (2)-(4), but 
keeping (1), we found that the key results of \cite{ajl2} remained
unaffected.\footnote{\label{footchange}However, one interesting change 
{\it did} occur:
in \cite{ajl2} we observed a first-order phase transition for large 
$\tilde k$ (there called $k_{0}$) to a phase where successive 
spatial slices decouple. We viewed the presence of this phase 
as a discretization artifact irrelevant for the continuum physics.
This interpretation is corroborated by our recent simulations
without the constraints (2)-(4), where the large-$\tilde k$ phase
is simply absent. It implies that the generalized model 
gives in some sense a better representation of continuum physics. 
This is reminiscent of 2d simplicial quantum 
gravity, where the continuum limit is approached much faster if
self-energy and tadpole graphs are allowed.\label{fnote}} 

The status of condition (1) is different. Recall that in the original 
representation one obtains the geometry at time
$t$ (at time $t+1$) from the quadrangulation at $t+1/2$ by shrinking all
dashed (solid) links to zero. The $A$- and $B$-graphs of the dual
picture at $t+1/2$ are of course individually precisely the
duals of these two-geometries at integer times. What does it
imply for the spatial quadrangulation at $t$ if the dual $A$-graph is
disconnected? Using the same ``shrinking-prescription'', 
it cannot separate into several 
pieces, but it does degenerate in the sense of forming 
a number of connected spherical graphs which touch each other 
pairwise only in single points. The resulting two-dimensional
space is therefore no longer a manifold, but a branched tree of
such spherical components. 

\begin{figure}[t]
\centerline{\scalebox{0.6}{\rotatebox{0}
{\includegraphics{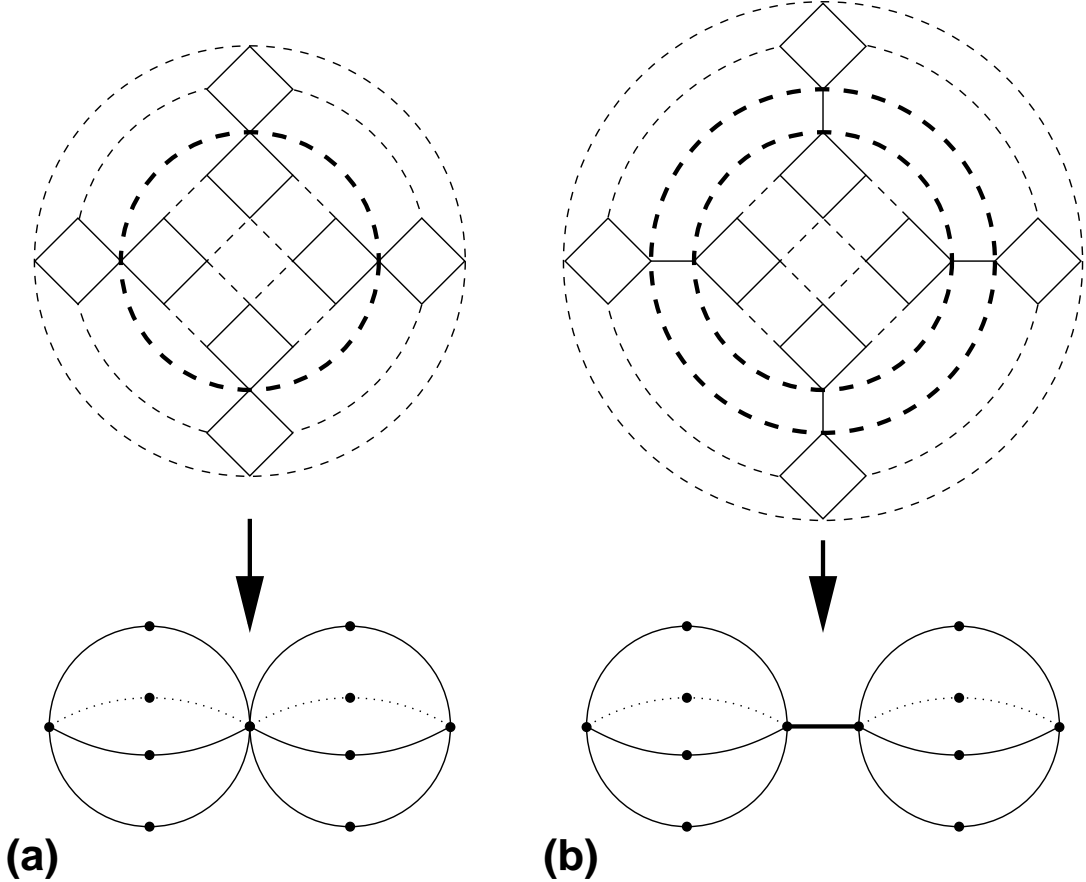}}}}
\caption[worms]{Examples of matrix-model configurations at $t+1/2$
which are not allowed in the original Lorentzian gravity model
and which result in geometries with wormholes at time $t$. 
Shrinking the dashed links to zero, one obtains the two-geometries 
at the bottom. The thick dashed lines in the quadrangulations at the 
top are contracted to
touching points or to points along one-dimensional wormholes.}
\label{worms}
\end{figure}
This phenomenon is
illustrated in Fig.\ \ref{worms}. The pictures at the top show two 
quadrangulations of a two-sphere at time $t+1/2$ made of the three
types of square-shaped building blocks. As two-dimensional 
configurations they look rather innocent and regular.  
However, they are pathological from a three-dimensional point of
view. Drawing the dual graphs, one finds in both cases 
that the $A$-subgraph, made of solid dual edges, consists of two 
components. Each component is dual to a quadrangulation of a 
two-sphere by four (solid-line) squares.
When shrinking away the dashed links of the original quadrangulation, 
the ``necks'' (indicated by the thick dashed circular lines) 
between the two spheres are gradually pinched to points. 
If there is just one neck, the two spheres will touch in a point
(Fig.\ \ref{worms}a).
If there are several concentric necks, with annuli of
dashed-solid squares (corresponding to closed rings of 
(2,2)-tetrahedra) in between, this process generates one-dimensional
``wormholes'', as illustrated in Fig.\ \ref{worms}b. 

Similar so-called ``touching''-interactions have been studied in the 
context of two-dimensional matrix models \cite{ddsw,kor}, and are closely
related to these wormholes (see Sec.\ \ref{reinter} for a detailed 
discussion). Although configurations of this type
are already present in ordinary matrix models of Euclidean 2d gravity,
the explicit introduction of such an interaction term in the action 
(and an associated coupling constant)
allows us to increase their weight and thereby control their 
abundance. Depending on the value of the coupling 
constant, one can obtain either the 
ordinary universal behaviour of two-dimensional Euclidean gravity, 
or a modified critical behaviour, described in more detail below.
We will see that in our 3d-gravitational reinterpretation of the
matrix model the presence of wormholes and touching-interactions 
is governed by the
value of the bare gravitational coupling constant $\tilde k$.

Let us finally note that the above discussion about degenerate
geometries could be repeated verbatim for the cubic matrix model with
partition function
\beq{pi3}
Z(\a_1,\a_2,\b) = \int \d A \d B \, 
\e^{-M \tr ( \oh A^2+\oh B^2 -\frac{\a_1}{3} A^3
-\frac{\a_2}{3} B^3 -\frac{\b}{2} ABAB)},
\eeq
which is associated with the original 3d simplicial gravity model
described in \cite{ajl1,d3d4} (with (3,1)- and (1,3)-tetrahedra 
instead of (4,1)- and (1,4)-pyramids).  
The only reason why we prefer to use the quartic model defined by \rf{2.9}
is that it {\it has} been solved for $\a_1\equ \a_2$ \cite{kz}. 
We would expect from 
universality arguments that the models given by \rf{2.9} and 
\rf{pi3} lead to the same continuum physics.

\section{Reinterpreting the matrix model}\label{reinter}

We have seen in the previous section that 3d Lorentzian ``sandwich'' 
geometries can be put into correspondence with a subclass of 
2d graphs generated by 
a quartic matrix model with $ABAB$-interaction. We have also argued that 
this generalization is potentially relevant, in the sense of the two 
models having a different phase structure. 

The phase structure of the
three-dimensional Lorentzian model has been investigated numerically
in \cite{ajl2,ajl3}. For given gravitational coupling $\tilde k$, there
is a critical value $\tilde\lambda_{c} (\tilde k)$ which is to be 
approached from {\it above}, that is, from the region $\tilde\lambda >
\tilde\lambda_{c}$ of the cosmological constant where the partition 
function converges. In the original simulations we also observed
a first-order transition for large $\tilde k$ when moving along the 
critical line $\tilde\lambda_{c} (\tilde k)$. As already mentioned in
footnote \ref{footchange}, this transition is a discretization 
artifact 
which disappears when one relaxes some of the manifold constraints.
What emerges as the phase structure for 3d Lorentzian gravity is
therefore simply that of a single phase (at least in the range of
coupling constants accessible to our computer simulations) 
where the taking of the 
continuum limit coincides with the tuning of the cosmological 
constant to its critical value, much as in two-dimensional models of 
quantum gravity. The role of the gravitational coupling $\tilde k$ is
merely to set an overall scale for the system, without affecting its
continuum properties, as we have argued in \cite{ajl2}. 

What we want to explore presently is the physical interpretation
in terms of three-geometry of the {\it generalized} model (where 
multiply-connected $A$- and $B$-graphs are allowed to occur as the 
spatial sections at half-integer times) and of its phase structure.
This is of course the two-matrix model defined by eq.\ (\ref{2.9}). 
Let us consider first the case of vanishing
boundary cosmological constants, $z_{i,f} \equ 0$. Although 
we will not be able to construct the Hamiltonian or the 
propagator (c.f. footnote \ref{footvanish}), we can nevertheless 
discuss the phase diagram of the model.
We will comment later on the general matrix model with $z_i,z_f \neq 0$.

\begin{figure}[t]
\centerline{\scalebox{0.6}{\rotatebox{0}
{\includegraphics{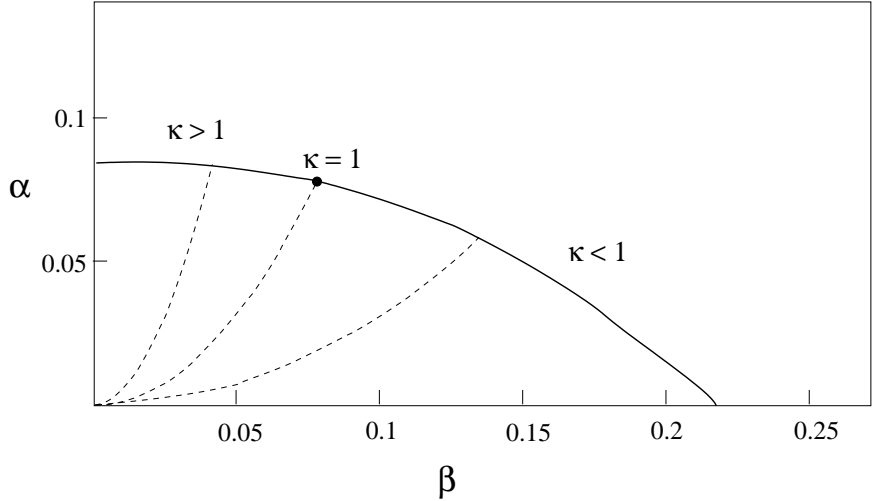}}}}
\caption[mphase]{The phase diagram of the $ABAB$-model, according to
\cite{kz}, with the critical line and the critical point at
$\a =\b =\frac{1}{4\pi}$ (corresponding to $\k =1$). 
One way of approaching the critical line
is through fine-tuning of the cosmological constant $\tilde\lambda$
along lines of constant $\tilde k$, shown as dashed curves. To end up
at the critical point, one should move along the line of constant
$\tilde k_c =-\frac{1}{3}\ln \b_c \simeq 0.85$.}
\label{mphase}
\end{figure}

Physically this choice of couplings implies that we are studying
the geometric fluctuations between 
two successive spatial slices with free boundary conditions.
As was clear from the computer simulations in \cite{ajl2},
the behaviour of this one-step system determines the phase structure 
of the discretized theory. Other than that, the choice 
$z_{i,f}\equ  0$ has the great advantage that the corresponding matrix 
model has already been solved. Details 
of the solution can be found in the paper of 
V.\ Kazakov and P.\ Zinn-Justin \cite{kz}.
Their analysis of the matrix model with $\a_1\equ \a_2 \equ \a$
can be summarized as follows. 
Let us fix a ratio $\k \equ \a/\b$ and gradually increase 
the values of $\a$ and $\b$ away from zero.
The large-$M$ limit of the matrix model is defined for
sufficiently small $\a$ and $\b$, and the model becomes critical at a
point $(\b_c(\k)$, $\a_c(\k)\equ \k \b_c(\k))$, giving rise to 
a {\it critical line} in the $(\b,\a)$-plane (see Fig.\ \ref{mphase}). 
Along it one finds two
phases, separated by a second-order phase transition at $\k \equ 1$,
with $\a_c(1) \equ \b_c(1)$. 
In the context of two-dimensional gravity, they were given the 
following interpretation \cite{kz}: the phase of $\k > 1$ with
$\a_c(\k) > \b_c(\k)$ can be viewed as 
ordinary two-dimensional Euclidean gravity, 
with central charge $c \equ 0$. In this phase 
the relation between the length $l$ of a ``typical'' loop of $A$-links 
(or $B$-links) and the area $\cal A$ of the two-dimensional 
surface enclosed by the loop is $l^2 \sim \cal A$. By contrast, in the 
other phase (which also has $c= 0$), the scaling is
anomalous, $l^{4/3} \sim \cal A$. 

One can understand these two situations by simply looking at the
two limits $\b \equ 0$ and $\a \equ 0$. 
By setting $\b \equ 0$ we switch off the $ABAB$-interaction,
leading to a decoupled system of two $\phi^4$ one-matrix models, whose
critical behaviour is individually that of an
ordinary 2d Euclidean gravity system. 
Setting $\a\equ 0$, we obtain a matrix model whose only interaction
comes from the term $\tr ABAB$. This model was first solved in \cite{ck}
by mapping it to the so-called dense loop phase of 
the $O(1)$-matrix model on random graphs.\footnote{ 
The $O(1)$-model on random graphs is known to correspond to $c \equ 0$ 
quantum gravity, but with two distinct phases 
\cite{kostov1,kostov2,ks}, not unlike those of the $ABAB$-model.
In a continuum interpretation, these result 
from assigning different boundary operators to the 
theory. The dilute loop phase corresponds to ordinary 
2d Euclidean quantum gravity where boundary lengths scale according 
to $l^2 \sim {\cal A}$, while in the dense phase one has
$l^{4/3} \sim \cal A$ \cite{kostov2,ks}.}

\subsection{Intermezzo: touching-interactions}\label{inter}

We will now reinterpret the results described above in the context of
three-dimen\-sio\-nal Lo\-rentz\-ian gravity. Recall our earlier discussion 
in Sec.\ \ref{matrix}, where we showed that from a three-dimensional 
point of view a generic feature of the matrix-model configurations
is the presence of touching points and wormholes in the associated
spatial slices. The matrix model therefore describes transitions
between two-geometries with a tree-like structure, consisting of
regular quadrangulations of two-spheres, pairwise connected by
one-dimensional wormholes of length $l\geq 0$ (Fig.\ \ref{worms}).

Let us for a moment revert back to a pure matrix-model language.
For the sake of definiteness, we will concentrate on a
quadrangulation at time $t$ made from solid squares.  
We will now show that the effective weight associated with the 
mutual touching points of the two-spheres making up the 
tree-configuration is obtained by integrating out the $B$-matrices.
While this cannot be done 
explicitly as long as the $\tr B^4$ term is present, it is instructive
to do the integration for the general matrix model (\ref{2.9})
with $\a_2 \equ 0$.\footnote{For $\a_{2}\not= 0$ there will be
other touching interactions beyond the ones described below.} 
In this case, the action is Gaussian in the $B$-matrix and we obtain
\bea
Z(\a_1,\a_2\equ 0,\b)& =& 
\int \d A \, \d B \; \e^{M \Big( -\oh \mathrm{tr}\, (A^2 +B^2) 
+\frac{\a_1}{4} \mathrm{tr}\, A^4 
+ \oh \b\ \tr ABAB\Big)}\nn 
&=& \int \d A \; \Big(\det (I - \b A^{\rm T} \otimes A)\Big)^{-\oh}\; 
\e^{ M \Big( -\oh \tr A^2  +\frac{\a_1}{4}  \tr A^4 \Big)}\nn
&=& \int \d A \; \; 
\e^{ M \Big( -\oh \tr A^2  +\frac{\a_1}{4}  \tr A^4  +\frac{1}{2M}
\sum_{k=1}^\infty 
\frac{\b^k}{k} \tr A^k \tr A^k \Big)}, \hfill \label{3.14}
\eea 
which now describes a particular one-matrix model.
Terms like $\tr A^k \tr A^k$ are usually referred to as 
``touching-interactions''. How can their effect be visualized in 
geometric terms? Firstly, in the same way as
$\tr A^{4}$ represents a square of solid edges, a term $\tr A^k$ 
in the action can be thought of as a $k$-gon. 
Expanding the action and performing Wick-contractions corresponds to 
gluing such polygons together by identifying their links pairwise.  
Imagine that during this process we had constructed a surface glued from 
various polygons, with a boundary consisting of $k$ links. 
By performing the appropriate Wick-contractions, we could now
glue a $k$-gon (represented by $\tr A^{k}$) to this boundary
and close off the surface. 

By extension, we can determine the effect of a term like $\tr A^k \tr A^k$.
Imagine a pair of surfaces made from polygons, both with a
boundary of length $k$. By contracting with $\tr A^k \tr A^k$, we can
close off the two surfaces. Because this happens
{\it simultaneously} for both of the surfaces, we associate by this 
process a $k$-gon of the first surface with another $k$-gon of the 
second surface, which we may think of as a touching point between
the two geometries (for example, located at the centres of the 
$k$-gons).  

\begin{figure}[t]
\centerline{\scalebox{0.5}{\rotatebox{0}{\includegraphics{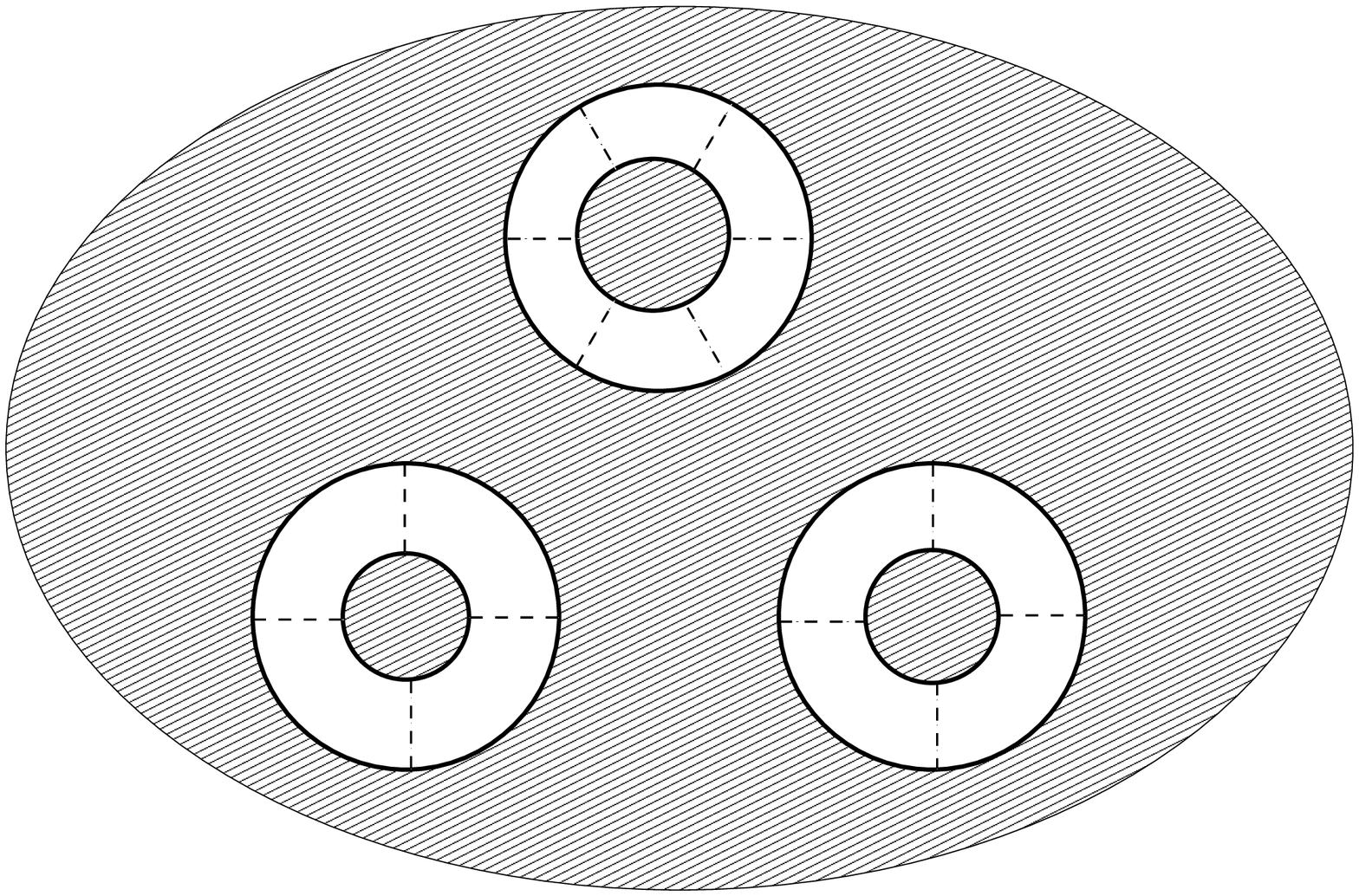}}}}
\caption[annuli]{A typical configuration generated by the matrix model
(\ref{2.9}) for $\alpha_{2}=0$. 
By integrating out the $B$-matrices one obtains touching terms
$\tr A^k \tr A^k$ (here, $k=4$ and 6). Their effect is to identify 
the boundary links of
the solid-square components pairwise across the ``gaps'' formed by
the solid-dashed squares, as described in the text.}
\label{annuli}
\end{figure}

This is illustrated by Fig.\ \ref{annuli}, which shows a typical 
quadrangulation at $t+1/2$ appearing in the matrix model with 
$\a_{2}=0$ (so no dashed squares are present), before performing the
$B$-integration. The shaded areas represent components made of solid 
squares, where the outer boundary should be thought of as a single 
point, since the entire geometry is spherical. 
Integrating out the $B$-matrices has precisely the effect of 
generating terms $\tr A^k \tr A^k$ and identifying the solid boundary 
links pairwise (each link with its partner lying opposite in a
solid-dashed-solid-dashed square). 
Within the one-matrix model it is of course a matter of convention how
one wants to think of the pairwise associations introduced by
the terms $\tr A^k \tr A^k$, whether as microscopic touching points
between the two closed-off surfaces or simply as regular gluings of
one boundary of length $k$ to another. However, 
it {\it is} natural 
to think of them as creating a {\it connection} between the surfaces.
There is a factor $1/M$ associated with each touching interaction,
and it is well-known that to leading order in $M$ one obtains 
through ``touching and connecting'' only two-geometries with an overall
spherical topology, and no spaces of higher genus. Collapsing all
the genuinely two-dimensional parts of this two-geometry to
one-dimensional lines, we would obtain a branched polymer, i.e.
a one-dimensional tree structure without any closed loops.

What is our motivation for this excursion into touching-interactions?
The general $ABAB$-model does not in itself suggest an
immediate physical interpretation in terms of spatial or space-time 
geometry. However, with the benefit of hindsight, one could have
taken the presence of the limiting case $\beta =0$ (where one has
two independent copies of Euclidean 2d gravity) as an indication
that switching on the $\beta$-coupling will lead to a theory of
two interacting two-geometries. The situation where such an interaction
occurs naturally is of course when the two-geometries appear
as neighbouring embedded spatial slices within a 3d space-time, with
their interaction dictated by the 3d Einstein action. Given the
three-dimensional interpretation, it is then natural to determine the
effect of the neighbouring two-geometry by integrating out the
corresponding degrees of freedom in the partition function (for 
instance, the $B$-matrices). As we have seen, this effectively
introduces touching-interactions in a 2d slice which are 
dictated by $\beta$, where $\b$ in turn depends on the coupling constants 
$\tilde\lambda$ and 
$\tilde k$ of three-dimensional quantum gravity. 

Another related point that follows from the three-dimensional
interpretation is that it gives us a {\it precise} geometric
way of obtaining a two-geometry at time $t$ (or $t+1$) from
a general matrix-model configuration at time $t+1/2$ by
the shrinking-prescription already mentioned in Sec.\ \ref{matrix}.
This may be visualized as a continuous process where the lengths of
either the dashed or the solid links are gradually shrunk to zero.
Although this prescription was originally invented to 
describe the well-behaved geometries of 3d Lorentzian quantum
gravity, it works just as well for the more general configurations
generated by the matrix model. We have already seen explicit
examples in Sec.\ \ref{model}, where this led to the creation of 
spatial wormholes (Fig.\ \ref{worms}). 

\begin{figure}[t]
\centerline{\scalebox{0.5}{\rotatebox{0}{\includegraphics{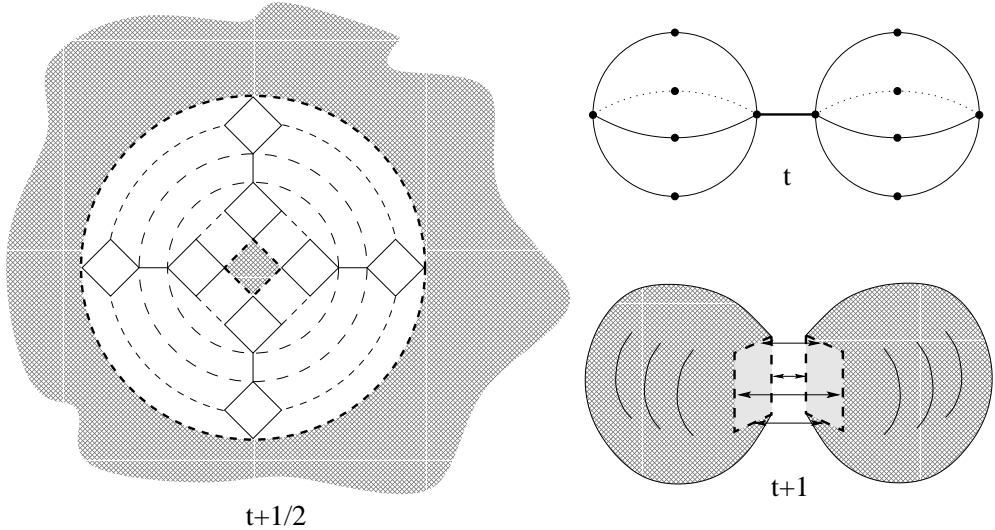}}}}
\caption[shrink]{A configuration of the $ABAB$-matrix model, 
giving rise to a two-geometry with a spatial wormhole at time $t$ and 
a two-geometry with ``touching-interaction'' at time $t+1$.
Viewed as part of a Lorentzian space-time, this configuration
is degenerate.}
\label{shrink}
\end{figure} 
What is important to notice
is that from a three-dimensional point of view, such wormholes
and the touching-interactions described in this subsection are
really two sides of the same coin. That is, taking a general
configuration of the $ABAB$-model which is not allowed in
the original Lorentzian model, one will find {\it both} types
of degeneracy when applying the shrinking-prescription.
We illustrate this with a simple matrix-model graph 
in Fig.\ \ref{shrink}. We start at time $t+1/2$ from an extended 
version of 
Fig.\ \ref{worms}b, where the shaded areas indicate regular
dashed quadrangulations. Since these will all vanish
when we shrink away the dashed edges, 
we obtain the same wormhole geometry at time $t$ as before. 
On the other hand, if we now shrink the solid edges to zero, we
are left with two pieces of dashed quadrangulations, each with a
boundary of dashed links of length four (the thick dashed
lines in Fig.\ \ref{shrink}). We then must identify the boundary
links pairwise, as indicated by the arrows. This is clearly
the same mechanism that was at play in the touching-interactions,
now generalized to a situation with $\a_2 \not= 0$. 
Although the resulting configuration at time $t+1$ 
is (unlike the wormhole at $t$) perfectly well-behaved as
a {\it two}-geometry, it is not acceptable
from our original three-dimensional point of view because of the way
it is ``embedded'' in the three-dimensional space-time.
We will in the following not distinguish between those two
types of geometric degeneracy and simply refer to them
as ``wormholes''.\footnote{This fits with the terminology
used in matrix models, where the touching points brought about by the 
touching-interactions are also sometimes thought of as ``wormholes''.}

\subsection{Phase structure in the presence of 
touching-interactions}\label{phase}

Having established the relation with one-matrix models containing
(generalized) touching-interactions, let us try to understand what
we can learn from the known phase structure of these models.
There are three phases when touching-interactions are added 
to the usual $\tr A^4$-interaction. In phase I the touching-interaction 
plays no role, and 
the critical behaviour is determined by the $\tr A^4$-interaction.
A typical configuration (with probability one in the 
scaling limit) is a connected, spherical quadrangulation of the type
that dominates Euclidean 2d gravity. 

In phase II, the touching- and the $\tr A^4$-interactions coexist
and compete. A typical two-geometry consists of many spherical 
2d ``baby''-universes of all sizes, which are connected to each other 
at touching points, thus forming a ``blown-up'' branched 
polymer\footnote{Further information about the properties 
of branched polymers can be found in \cite{branched}.}. 
As was first realized in \cite{durhuus}, the critical behaviour results 
from an interplay between the properties Euclidean gravity and those 
of branched polymers (see also \cite{adj}). 

Quite surprisingly, {\it all} critical aspects of 
this phase can be understood from continuum 
quantum Liouville theory (which is usually thought to describe 
only phase I) \cite{klebanov1,klebanov2,klebanov3}. 
Let us review this briefly. Consider a conformal field theory in 
dimension two with a fixed background metric $\hat{g}$, and let $\Phi$ be 
a spinless primary field with scaling dimension $\Delta^{(0)}_\Phi$. 
Then the ``one-point'' function for $\Phi$ scales as  
\beq{conf1}
F^{(0)}_\Phi({\cal A})= \left\langle 
\int \d^2 \xi \, \sqrt{\hat{g}(\xi)} \,\Phi(\xi) \right\rangle_{CFT}
\sim {\cal A}^{1-\Delta^{(0)}_\Phi},
\eeq
where ${\cal A} = \int_{\Sigma} \d^2 \xi \, \sqrt{\hat{g}(\xi)}$ is the area
of the underlying manifold $\Sigma$, and the expectation value is taken in
the conformal field theory. Coupling this
theory to 2d Euclidean quantum gravity, the metric $g$ will be allowed
to fluctuate. We can decompose a general metric $g$ into 
$g \equ e^\phi \hat{g}$, where $\phi(\xi)$ is 
the conformal field. In conformal gauge, the integration 
over the fluctuating metric becomes an integration over the 
conformal factor $\phi$, weighted by the Liouville action $S_{L}[\phi]$,
while the spatial integral appearing in the one-point function is 
changed according to 
\beq{conf2}
\int \d^2 \xi \sqrt{\hat{g}(\xi)}\,  \Phi(\xi) \longrightarrow  
\int \d^2 \xi \sqrt{\hat{g}(\xi)}  \,\e^{\b_\Phi \phi(\xi)} \Phi(\xi). 
\eeq
The dressing exponent $\b_\Phi$ is determined by requiring that the 
``dressed'' operator $e^{\b_\Phi \phi}\Phi$ have conformal dimension
dimension (1,1), so that it can be integrated over the 
two-dimensional surface without breaking conformal invariance. This 
leads to the relation
\beq{conf3}
\b_\Phi (\b_\Phi +Q)= -2+2\Delta_\Phi^{(0)},~~~~Q= \sqrt{\frac{25-c}{3}}.
\eeq
The special case of the unit operator $\Phi = {\bf 1}$ corresponds to 
$\Delta_{\bf 1}^{(0)} \equ 0$, with dressing exponent $\b_{\bf 1}$.
The one-point function for the gravity-coupled theory is defined by
\bea
F_\Phi({\cal A}) &=& 
\left\langle 
\int \d^2 \xi \, \sqrt{\hat{g}(\xi)} \, \e^{\b_\Phi \, \phi(\xi)}
\Phi(\xi) \right\rangle_{QG} \nn
&:=&\frac{\int \cD \phi\; \e^{\frac{c-25}{48\pi^2} S_{L}[\phi]} \;
\del(\int \d^2 \xi \sqrt{\hat{g}}\,\e^{\b_{\bf 1} \phi}\mi {\cal A}) 
\left\langle 
\int \d^2 \xi \, \sqrt{\hat{g}} \,\e^{\b_\Phi \, \phi}\,
\Phi \right\rangle_{CFT}}{Z({\cal A})},
\label{conf4}
\eea
where $Z({\cal A})$ is the partition function with fixed area $\cal A$ 
(enforced by including the same $\del$-function that appears in the
numerator of \rf{conf4}).
By requiring that $F_\Phi({\cal A})$ be independent of the fiducial  
background metric $\hat{g}$, a
change of integration variables $\phi \to \phi + \frac{1}{\b_{\bf 1}}\log 
{\cal A}$ 
leads to
\beq{conf5}
F_\Phi ({\cal A}) = {\cal A}^{\b_\Phi/\b_{\bf 1}} \; F_\Phi(1).
\eeq
In analogy with \rf{conf1}, we now {\it define} the critical exponent 
$\Delta_\Phi$ of $\Phi$ for the gravity-coupled theory by
\beq{conf6}
F_\Phi({\cal A}) \sim {\cal A}^{1-\Delta_\Phi}.
\eeq 
Combining this with relation \rf{conf5}, one derives
\beq{conf7a}
\Delta_\Phi = 1-\frac{\b_\Phi}{\b_{\bf 1}}.
\eeq

For our present purposes it is important to note that eq.\ \rf{conf3} 
has {\it two} solutions, namely,
\beq{conf7}
\b_{\Phi}^{(\pm )} = 
\frac{-(25-c) \pm \sqrt{(25-c)(1-c + 24\Delta^{(0)}_\Phi)}}{6Q}.
\eeq 
In conventional quantum Liouville theory one works with 
$\b_{\Phi}^{(+)}$ because this choice ensures that in the ``classical'' 
limit $c=-\infty$ (classical in the sense that the fluctuations of the 
Liouville field in \rf{conf4} are completely surpressed) the
scaling dimensions of the model without gravity coupling are
recovered,
\beq{conf8}
\Delta_\Phi \to \Delta_\Phi^{(0)}~~~{\rm for}~~ c \to -\infty,
\eeq
as can easily be verified from formulas \rf{conf7} and \rf{conf7a}.

It is a remarkable fact that all critical exponents 
which can be calculated in phase II of the matrix model with 
touching-interactions 
are obtained by replacing $\b_\Phi^{(+)} \to \b_{\Phi}^{(-)}$,
changing the scaling relations according to 
\beq{conf5a}
F_\Phi^{(+)} ({\cal A}) \sim {\cal A}^{\b_\Phi^{(+)}/\b_{\bf 1}^{(+)}} 
\longrightarrow 
F_\Phi^{(-)} ({\cal A}) \sim {\cal A}^{\b_\Phi^{(-)}/\b_{\bf 1}^{(+)}}.
\eeq
The implications of this prescription can be illustrated 
by considering the simplest boundary operator,
\beq{conf9}
\ell = \oint_{\partial\Sigma} \d s \sqrt{\hat{e}(s)}, 
\eeq
which measures the length of the boundary of the 
two-dimensional manifold $\Sigma$ as a function of the induced metric 
$\hat{e}(s)$ on $\partial\Sigma$. 
Unlike in matrix models, it is somewhat awkward to introduce
boundary operators like \rf{conf9} in Liouville theory. 
Nevertheless, $\ell$ can be treated along similar lines 
as the operator $\Phi$ in \rf{conf2} above. The expression one obtains 
for the gravitationally dressed version of the length operator is
\beq{cone10}
\ell= \oint \d s \sqrt{\hat{e}(s)}\, 
\e^{\b_\ell^{(+)} \phi(s)}, ~~~~\b_\ell^{(+)} = \oh \b_{\bf 1}^{(+)}.
\eeq
From the point of view of conformal field theory it is rather surprising
that the ``naive'' value $\b_\ell \equ \oh \b_{\bf 1}$ (i.e. the boundary 
scales like the square root of the bulk) is correct. 
However, as can be read off from \rf{conf5}, this does lead to the 
canonical scaling 
\beq{conf11}
\la \ell^{(+)} ({\cal A})\ra_{QG} \sim {\cal 
A}^{\b_\ell^{(+)}/\b_{\bf 1}} 
=  {\cal A}^{1/2}
\eeq
already known from matrix-model considerations.

To determine the scaling relevant in phase II, we should now
replace the operator $\b_\ell^{(+)}$. It turns out that the correct
substitution for this particular boundary operator is given by
\beq{repla}
\b_\ell^{(+)} \longrightarrow \b_\ell^{(-)} = \oh \b_{\bf 1}^{(-)}.
\eeq
This leads to the scaling behaviour
\beq{conf12}
\la \ell^{(-)} ({\cal A})\ra_{QG} \sim 
{\cal A}^{\b_\ell^{(-)}/\b_{\bf 1}^{(+)}} =  {\cal A}^{3/4},
\eeq  
implying that the boundary scales anomalously in terms of
the area. This anomalous scaling is clearly a reflection
of the additional fractal structure introduced by the 
touching-interactions.
Note that it coincides with the spectral dimension of 
branched polymers \cite{spectral1,spectral2}.
  
The scaling \rf{conf12} coincides with the scaling behaviour observed 
in the $ABAB$-matrix model in the phase with $\kappa < 1$, and
mentioned already earlier in Sec.\ \ref{reinter}. 
This corroborates our interpretation of this model in terms
of interacting two-geometries.
Furthermore, it agrees with the relation between loop length and area
found in the dense phase of the $O(1)$-model, where the replacement 
\rf{repla} was first noticed \cite{ks}.

Lastly, the critical behaviour in phase III of the matrix model with 
touching-interactions is characterized by a complete dominance of
branched polymers. The size of the individual spherical components
making up the 2d universe never exceeds the cutoff scale. This
phase does not seem to have an analogue in the $ABAB$-matrix model.

\subsection{Summary of Sec.\ \ref{reinter} }\label{summary} 

\begin{figure}[t]
\centerline{\scalebox{0.6}{\rotatebox{0}
{\includegraphics{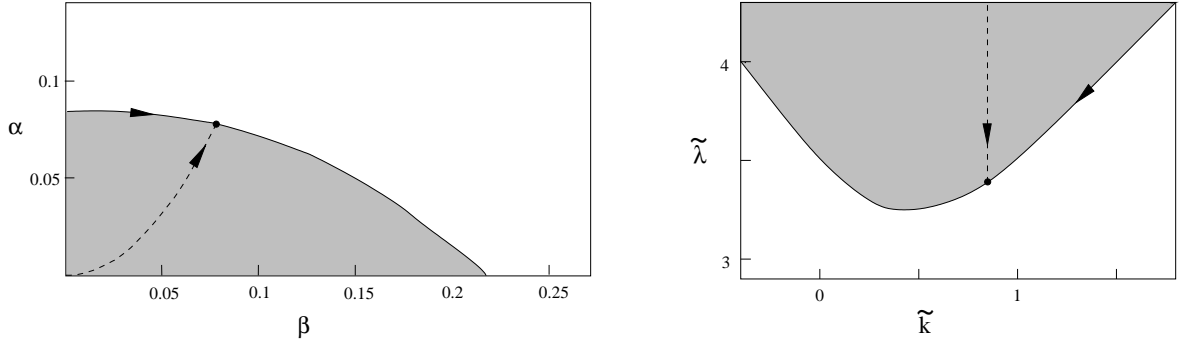}}}}
\caption[bothphase]{Translating the phase diagram of the $ABAB$-matrix 
model in terms of the original coupling constants $\a$ and $\b$ (left) 
to the couplings $\tilde k$ and $\tilde\lambda$ associated with
its interpretation in terms of 3d Lorentzian gravity (right).}
\label{bothphase}
\end{figure}

Let us summarize what we have learned about the
three-dimensional geometric interpretation of the $ABAB$-model. To this 
end, we will first translate its phase diagram (for $\a_{1}=\a_{2}$) 
to the $\tilde k$-$\tilde\lambda$-plane of the gravitational couplings. 
We deduce from \rf{2.12} that
\beq{3.15}
\frac{\a}{\b^2} = \e^{3\tilde k} ~~~~{\rm and}~~~~
\frac{1}{G_N} \sim \tilde k = \frac{1}{3} \log \frac{\k}{\b_c(\k)},
\eeq
where the second relation holds along the critical line (the points
where a non-trivial continuum limit exists). The shaded areas in
the two phase diagrams of Fig.\ \ref{bothphase} indicate the region
in which the partition function converges. The dashed curves in
both diagrams show an approach to the critical line along a curve
of constant $\tilde k$. Moving from left to right 
along the critical curve of the left phase diagram (so that $\b$ 
increases and $\k$ decreases) translates into a motion from right
to left along the critical line of the ``gravitational'' phase diagram 
(that is, towards a
smaller $\tilde k$ or a larger bare Newton's constant $G_{N}$). 

In terms of space-time geometry, at the point $G_{N}=0$ (corresponding 
to $\b =0$) subsequent spatial slices 
are completely decoupled. No information can be propagated in time and
the system has no interesting three-dimensional properties. 
Increasing $G_{N}$ away from zero, we enter the phase $\k >1$ of the
matrix model. Interactions between neighbouring 
spatial slices of the 3d universe become possible\footnote{We 
cannot be more specific about the nature of the interactions
before having analyzed the transfer matrix of the model in more 
detail.}, and the geometrical properties of a typical spatial slice at 
integer-$t$ resemble those generated in 2d Euclidean quantum gravity. 
It is very suggestive to identify this phase with that of Lorentzian 
gravity defined in \cite{ajl1,d3d4}, since the simulations of
\cite{ajl2,ajl3}
revealed both the existence of extended three-dimensional space-times
(indicative of correlations in time-direction), 
and evidence that the Hausdorff dimension of the spatial slices 
is $d_{H}=4$, in agreement with that of 2d Euclidean gravity. 

Apparently, the phase with $\k\leq 1$ cannot be realized in the
Lorentzian manifold model because it does not allow for the
creation of ``wormholes''\footnote{At least, we have so far not seen 
any evidence of this phase or the associated second-order
transition in the computer simulations.}. 
By contrast, no such restriction
exists in the $ABAB$-matrix model, and by gradually increasing
Newton's constant $G_{N}$ such wormhole configurations become
energetically more favoured. Looking at spatial slices in
the phase $\k\leq 1$, a typical two-geometry will consist of
many smaller universes connected to each other by cutoff-size wormholes.

\section{The general ABAB-matrix model}\label{general}

While the phase structure of the matrix model of (generalized) 
3d Lorentzian gravity can be understood from the KZ-solution for
the special case $\a_1\equ \a_2$ of the $ABAB$-matrix model, 
the construction of the 
quantum Hamiltonian requires that we perturb away from $\a_1\equ \a_2$.
It is not difficult to generalize the ansatz of \cite{kz}
to this situation, using the techniques developed in
\cite{ksw}. One can write down a set of singular integral 
equations which generalize those solved in \cite{kz}. 
Furthermore, using results from the general theory
of singular integral equations one can prove that the solution is
unique. 

Although we have at this stage nothing much to say about the
explicit solution for general $\a_1$ and $\a_2$,
it may be worth pointing out a connection with yet another
2d statistical model. As mentioned
in \cite{kz} and studied in more detail in \cite{zj},
the $ABAB$-matrix model with $\a_1\equ \a_2$
can be mapped to an eight-vertex model 
defined on a random four-valent lattice (in the dual picture).
The map is simply given by $X\equ A+iB$. 
Using the same transformation, one can map
the general $ABAB$-model to a {\it sixteen}-vertex model
with partition function 
\beqv
Z(\a_1,\a_2,\b)\equ \!
\int \!\!\d X \d X^\dg \,
\e^{-N\tr \Big[ \oh X^\dg X - b X^2{X^\dg}^2 -\frac{c}{2}(XX^\dg)^2
-\frac{d}{4} (X^4+{X^\dg}^4)- f (X^3X^\dg+X {X^\dg}^3)\Big]} ,
\eeq
where the constants {\it b, c, d,} and $f$ are given by
\beq{vertex1}
b= \frac{\a_1+\a_2+2\b}{16},~~~~ c=d= \frac{\a_1+\a_2-2\b}{16},
~~~~f= \frac{\a_1-\a_2}{16}.
\eeq
Since the eight-vertex model on a regular square lattice 
has already been solved, it may not be very surprising that the 
restricted
$ABAB$-model can be solved too. This would be in line with
the general observation that it is often simpler to solve
matter models on random dynamical lattices (associated with
2d quantum gravity) rather than on regular ones. 
By the same token, one should maybe not be too discouraged by the 
fact that the {\it general} sixteen-vertex model has not been solved 
on a regular lattice. This is also not what is needed here, because
of the particular form of the parameters \rf{vertex1}.

\section{Taking the continuum limit}\label{cont}

We have already described in some detail in Subsection \ref{summary}
how the phase structure of the matrix model with $ABAB$-interaction 
can be reinterpreted from the point of view of three-dimensional geometry, 
and how this fits into our previous investigations of three-dimensional 
Lorentzian quantum gravity. One further point we would like to make
concerns the nature of the continuum limit in the gravitational
model, and what the newly established relation with the matrix model 
can add to our understanding of it.

In dynamically triangulated models of quantum gravity 
one usually performs the continuum limit by fixing the
bare (inverse) gravitational coupling constant $\tilde k$
and fine-tuning the bare cosmological constant $\tilde\lambda$ to its 
critical value $\tilde\lambda_{c}(\tilde k)$. Since the space-time volume is 
conjugate to the cosmological constant in the action, such a fine-tuning 
corresponds to taking the lattice volume $N_{tot} \equ 
N_{14}+N_{41}+\oh N_{22}$ to infinity, and 
can be viewed as an additive renormalization 
\beq{6.1}
\tilde\lambda = \tilde\lambda_{c}(\tilde k) + \Delta \tilde\lambda,
~~~~\Delta \tilde\lambda = \L_3 a^\nu,
\eeq
of the cosmological constant, where $\Delta \tilde\lambda$ should 
be related to the continuum cosmological 
coupling constant $\L_{3}$ by a suitable scaling. 

The analogue of this procedure in a lattice field theory, for example, 
the Ising model defined on a hypercubic $N^{d}$-lattice,
would be to take the infinite-lattice limit. However, this may not
lead to a {\it continuum} limit. In the case of the 
Ising model, in order to obtain such a limit, the coupling constant $\b$ 
(the inverse temperature) must always be fine-tuned to a critical value, 
even if the lattice volume was infinite.
By analogy, one might therefore expect that also
in simplicial gravity models, the gravitational coupling $\tilde k$
had to be fine-tuned to a critical value $\tilde k_c$ to 
arrive at an interesting continuum theory. 

However, this is not the only possible scenario. Consider another
hypercubic lattice model in $d$ dimensions, one
with a lattice scalar field $\phi_n$ and the simple Gaussian action
\beq{6.2}
S[\phi] = \sum_n \sum_{i=1}^d(\phi_{n+\hat{e}_i} -\phi_n)^2.
\eeq
For this model, one {\it automatically} obtains the continuum Gaussian 
field theory by taking the limit of infinite lattice volume and rescaling 
the lattice spacing to zero. The same situation is encountered in
two-dimensional simplical quantum gravity, both for Lorentzian and
Euclidean signature, when formulated as a
sum over triangulations with geodesic edge lengths determined by 
a single lattice spacing $a$. 
Fine-tuning the cosmological coupling constant $\tilde\lambda^{(2)}$
to its critical value according to 
\beq{6.3}
\tilde\lambda^{(2)} = \tilde\lambda^{(2)}_{c} + \Delta \tilde\lambda^{(2)},
~~~~\Delta \tilde\lambda^{(2)} = \L_2 a^2,
\eeq
or, equivalently, taking the infinite-volume limit 
and scaling $a$ to zero, leads automatically 
to a continuum theory of 2d quantum gravity. In \rf{6.3}, 
$\L_2$ denotes the continuum two-dimensional cosmological constant,
which has already appeared in our earlier discussion of Lorentzian 
2d gravity (see eq.\ \rf{2d.2}).

It is easy to see that the taking of the continuum limit in the
two-matrix model with $ABAB$-interaction falls into this latter
category. According to \cite{kz}, the continuum limit is obtained 
by fixing the ratio $\k\equ \a/\b$ (corresponding to straight lines
through the origin in Fig.\ \ref{mphase}), and increasing $\a$ and $\b$ until 
the critical point is reached. This is a natural procedure to adopt, 
since the ratio $\a/\b$ appears in the equations which determine the 
solution of the model. 
However, one could equally well approach 
the critical curve $\a_c(\k) = \k \b_c(\k)$ along any family of curves 
in the $(\b,\a)$-plane which intersect this line.
In particular, one could choose curves of constant $\tilde k$, which
according to \rf{3.15} are defined by $\a = \tilde k \b^2$ (c.f. 
Fig.\ \ref{mphase}). The curve for a given value $\tilde k$ intersects 
the critical line in the point $(\b_c(\k(\tilde k)), \a_c(\k(\tilde k)))$.
Moving along the constant-$\tilde k$ curve to the critical point
is of course tantamount to fine-tuning the cosmological constant, 
described by eq.\ \rf{6.1}. 
 
This observation lends further support to our interpretation
of the role of the gravitational coupling constant $\tilde k$ in
3d Lorentzian quantum gravity. Based on the results of our computer 
simulations we suggested in \cite{ajl2} that the continuum limit of 
this theory does {\it not} involve a renormalization of $\tilde k$. 
Rather, $\tilde k$ should act as a finite overall scale parameter.
Now, as we have seen in Sec.\ \ref{reinter}, it is natural to
identify the Lorentzian model of \cite{ajl1,d3d4} with the
$\kappa > 1$ phase of the $ABAB$-matrix model. 
Forgetting for a moment about the new three-dimensional interpretation
we have given it, the matrix model by itself is simply a 
two-dimensional statistical system with two phases and a transition 
point in between. In one of the phases, the scaling behaviour
of the subgraphs of $A$- or $B$-type corresponds 
to that found in ordinary 2d Euclidean quantum gravity with $c \equ 0$. 
The other phase corresponds also to $c \equ 0$ quantum gravity,
but with the ``unconventional'' branch of the KPZ equation
chosen in the way described in subsection \ref{phase}. 
Changing $\k$ (that is, changing $\tilde k$), but staying within the
same phase, the continuum limit is of course identical, which means
that the theories can differ at most by the choice of an overall 
length scale.

\section{Outlook}\label{outlook}

As we have already mentioned earlier, extracting further
physical information about 3d Lorentzian gravity, with or
without wormholes, requires a more detailed understanding 
of the transfer matrix (or, equivalently, of the quantum Hamiltonian).
This should help us to determine the appropriate renormalization
of the boundary cosmological constants in the continuum limit. 
We have seen in Sec.\ \ref{hamil} that in the two-dimensional Lorentzian 
model, where the transfer matrix, the Hamiltonian and their spectra are 
known explicitly, the analogous scaling is canonical.

Another interesting question is how the presence of the
extended three-dimen\-sio\-nal structures observed in simulations
of the Lorentzian model is reflected in the mathematical
properties of the transfer matrix. We found in \cite{ajl2}
that when the extension in time-direction of the simulated
three-geometries exceeds $\L_{3}^{-1/3}$ -- with the
renormalized cosmological constant $\L_{3}$ setting an
effective volume scale of the system -- a universe with a
finite spatial volume exists only for a time duration
$T \sim \L^{-1/3}$, whereas during the remaining time 
the size of the (compact) spatial directions remains at the
cutoff scale. 

Our interpretation of the $ABAB$-matrix model raises the interesting 
issue of whether the phase transition of this model can be given
a meaning in terms of three-dimensional geometry too. 
The intriguing fact that a two-dimensional central charge $c \equ 1$
is associated with this point brings to mind attempts of associating
2d conformal field theories with
two-dimensional (boundary) surfaces in theories of 3d gravity.
Examples are given by the recent entropy calculations in anti-de Sitter 
\cite{ads} and de Sitter space \cite{ds}. We cannot directly
compare with any of these results, because our set-up is quite
different. We are working in a geometric and not a gauge-theoretic
Chern-Simons formulation, our boundary surfaces are compact and 
space-like,
and our (bare) cosmological coupling constant is by necessity
positive. Nevertheless, the $ABAB$-matrix model is suggestive of the
presence of additional symmetries at the phase transition point.

As for the topology of our model, we have restricted ourselves
to discussing space-times $S^2\times [0,1]$, since the
solution of the $ABAB$-matrix model as presented in \cite{kz} 
is valid only for spherical topology. In order to make a
comparison with other approaches to (2+1) quantum gravity,
it would be desirable to consider also surfaces of higher genus $g$,
where the physical configuration space is described by a
finite number of Teichm\"uller parameters. 
Results obtained so far for the matrix model 
suggest that its phase structure does not change for
higher genus; the values of the critical coupling constants are 
independent of $g$, and most critical exponents are 
unaltered since they refer only to the short-distance behaviour
of the model. Nevertheless it will be difficult to compare
with a canonical reduced phase space quantization, say, 
because in the matrix model all modular parameters appear 
{\it integrated over} for a given genus. It is possible one
could find a way to probe the individual Teichm\"uller parameters,
although it seems already quite complicated to perform the necessary 
large-$M$ expansion of the $ABAB$-matrix model
using the character expansion of \cite{kz}.

\vspace{12pt}

We have argued in this article that under certain assumptions 
the two-matrix model with $ABAB$-interaction 
describes three-dimensional Lorentzian quantum gra\-vi\-ty. 
In fact, not only does
it describe the ``regular'' Lorentzian quantum gravity
formulated in \cite{ajl1,ajl2}, but a more general theory
where space can split into many components connected 
by ``wormholes'', resulting in a tree structure of spatial universes 
forming a branched polymer. Starting at
vanishing bare Newton's constant $G_N$, by increasing $G_N$ gradually
we first find a phase of ``weak gravity'' which seems to
coincide with the phase of regular Lorentzian gra\-vi\-ty seen
previously in computer simulations. Eventually, we meet a
second-order transition point beyond which lies a phase
of ``strong gravity'' and large $G_N$. From our original point of 
view of using causality as an effective regulator of quantum
geometry, it is of course unclear to what extent the 
space-times with an abundance of spatial wormholes found in
the strong-gravity phase are still acceptable in the path integral.
In order to decide this question, we need to get a better
understanding of the genuinely {\it three}-dimensional
properties of this system of quantum geometry.

\subsection*{Acknowledgements} All authors
acknowledge support by the
EU network on ``Discrete Random Geometry'', grant HPRN-CT-1999-00161, 
and by ESF network no.82 on ``Geometry and Disorder''.
In addition, J.A. and J.J. were supported by ``MaPhySto'', 
the Center of Mathematical Physics 
and Stochastics, financed by the 
National Danish Research Foundation, and
J.J. by KBN grant 2P03B 019\,17.

\section*{Appendix}

In this appendix we will calculate the Regge version of the
Euclidean gravitational action \rf{2.1}, which we use in
Sec.\ \ref{model}. It should be remembered that our
geometries are Wick-rotated versions of discrete Lorentzian
(2+1)-dimensional space-times \cite{d3d4}. For a change, we will
do the angle calculations for Euclidean signature, and rotate
back afterwards. Let the ratio of the squared lengths of time- and
space-like links be given by a constant\footnote{The constants 
$\alpha$ and $\beta$ used in this appendix have nothing to do with
the couplings $\alpha$ and $\beta$ appearing in the main text.} 
$\b > \oh$, such that $l_{t}^{2}=\beta l_{s}^{2}$.
The dihedral angles of (4,1)- and (1,4)-pyramids around
space- and time-like links will be denoted by $\th_s$ and $\th_t$
respectively. Calling the corresponding 
dihedral angles of the (2,2)-tetrahedra $\phi_s$ and $\phi_t$,
one derives the relations
\beq{A2.2a}
\cos^2 \th_s = -\cos \th_t = \frac{1}{4\b -1},~~~~~~
\phi_s = \pi \mi 2\th_s,~~~\phi_t = \pi \mi \th_t.
\eeq 
The numbers of the three different types of building blocks
filling out the space-time between $t$ and $t+1$ are
$N_{14}(t)$, $N_{41}(t)$ and $N_{22}(t)$. From this we 
compute the number of time-like links between $t$ and $t\pl 1$ as 
$N_{1}^{\rm TL}(t) =N_{14}(t)\pl N_{41}(t) \pl N_{22}(t)\pl 2$. 
The number of spatial links contained in the spatial slice at $t$ 
is $2N_{41}(t)$, and 
each spatial link at $t$ belongs to two (4,1)-pyramids. Collecting 
all this information and using the well-known expressions
for Regge curvature in terms of dihedral angles \cite{regge}, including
boundary terms \cite{haso}, the curvature contribution to 
the total action is found to be 
\bea
\lefteqn{\frac{1}{2}\int_{M} \d^3 x \sqrt{g}\, R(x) + 
\int_{\partial M} 
\d^2x \sqrt{h}\, K(x) \longrightarrow} &&\label{A2.3}\\
&&\sum_{t} a \Big(4\pi \sqrt{\b}+
\Big[ (2\pi - 4\th_t)\sqrt{\b}+(2\pi -4\th_s) \Big]
(N_{41}(t)+N_{14}(t)-N_{22}(t))\Big), \no
\eea
where the term in square brackets is always positive. The total volume 
contributing to the cosmological term in the action is given by
\beq{A2.4}
\int_{M} \d^3 x \sqrt{g}\; \longrightarrow\; 
\frac{ a^3\sqrt{\b-\oh} }{3} \,
\sum_{t} \Big( 
N_{41}(t)\pl  N_{14}(t) \pl \oh N_{22}(t)\Big). 
\eeq
Putting in the appropriate (bare) coupling constants, this leads
to a total discretized action
\beq{A2.5}
S[\tilde k,\tilde\lambda] = c_0 t -\tilde k 
\Big(N_{41}+ N_{14}- N_{22}\Big)+
\tilde\lambda \Big(N_{41}+ N_{14}+  \oh N_{22}\Big),
\eeq
where now the total numbers of the different building
blocks appear and where the constants are given by
\begin{eqnarray}
&&\tilde k =\frac{a}{4\pi G_N}\Big(-\pi\sqrt{\beta} +2\sqrt{\beta}
\,\arccos\frac{1}{4\beta -1} +\arcsin \frac{\sqrt{\beta -\frac{1}{2}}}{
\beta-\frac{1}{4}} \Big),
\nonumber\\
&&\hspace{1cm}\tilde\lambda=\frac{a^{3}\Lambda}{24\pi G_N} 
\sqrt{\beta -\frac{1}{2}},
\;\;\;\;
c_0 = -\frac{a \sqrt{\beta}}{2 G_N}.
\label{A2.6}
\end{eqnarray}
In order to obtain the Lorentzian version of the action
(\ref{A2.5}) (up to an imaginary factor $-i$), we need to
replace $\b \to -\a$ such 
that $l_t^2 = -\a l_s^2$, in accordance with the notation 
in \cite{ajl1,d3d4}. To get the correct Lorentzian action,
one then inverts the prescription given in \cite{ajl1,d3d4}
by continuing the square-root expressions according to
$\sqrt{\b} \to i \sqrt{\a}$ and
$\sqrt{\b-\oh}\to i \sqrt{\a+\oh}$.

\end{document}